\documentclass{JHEP3}

\usepackage{amsmath}
\usepackage{amssymb}
\usepackage{latexsym}
\usepackage{graphicx}
\usepackage{psfrag,fancyhdr,epsfig}

\newcommand{\bsig}{\mbox{\boldmath $\sigma$}}
\newcommand{\evec}{\mbox{\boldmath $\xi$}}
\newcommand{\bM}{\mathbf{M}}
\newcommand{\bN}{\mathbf{N}}
\newcommand{\bP}{\mathbf{P}}
\newcommand{\nn}{\nonumber}
\newcommand{\Ain}{A^\text{in}}
\newcommand{\Aout}{A^\text{out}}
\newcommand{\pardiff}[2]{\frac{\partial#1}{\partial#2}}
\newcommand{\parddiff}[2]{\frac{\partial^2#1}{\partial#2^2}}
\newcommand{\pparddiff}[3]{\frac{\partial^2#1}{\partial#2\,\partial#3}}
\newcommand{\Pplus}{P_+}
\newcommand{\Pminus}{P_-}

\title{Brane decay of a (4+n)-dimensional rotating\\[1mm] black hole. III:
spin-1/2 particles}

\author{Marc Casals and Sam Dolan\\
School of Mathematical Sciences, University College Dublin,
Belfield, Dublin 4, Ireland. \\
E-mail: \email{Marc.Casals@ucd.ie}, \email{Sam.Dolan@ucd.ie}}

\author{Panagiota Kanti\\
Department of Mathematical Sciences, University of Durham,
Science Site, South Road, Durham DH1 3LE, United Kingdom.\\
E-mail: \email{Panagiota.Kanti@durham.ac.uk}}

\author{Elizabeth Winstanley\\
Department of Applied Mathematics, The University of Sheffield,
Hicks Building,\\ Hounsfield Road, Sheffield S3 7RH, United
Kingdom.
\\ E-mail: \email{E.Winstanley@sheffield.ac.uk}}

\abstract{ In this work, we have continued the study of the Hawking radiation
on the brane from a higher-dimensional rotating black hole by investigating
the emission of fermionic modes. A comprehensive analysis is performed
that leads to the particle, power and angular momentum emission rates, and
sheds light on their dependence on fundamental parameters of the theory,
such as the spacetime dimension and angular momentum of the black hole.
In addition, the angular distribution of the emitted modes, in terms of
the number of particles and energy, is thoroughly studied. Our results are
valid for arbitrary values of the energy of the emitted particles,
dimension of spacetime and angular momentum of the black hole,
and complement previous results on the emission of brane-localised
scalars and gauge bosons.}


\preprint{}

\begin{document}

\newpage
\section{Introduction}

The formulation, a few years ago, of theories \cite{ADD, RS} postulating the
existence of additional spacelike dimensions in nature have radically changed
the landscape in gravitational physics (for some early works towards this
direction, see \cite{early}). Driven by the motivation to address the hierarchy
problem, these theories leave unchanged the usual 4-dimensional particle physics
by localising the Standard Model (SM) particles on a 4-dimensional brane while
allowing the gravitons to propagate in the bulk -- the whole of the
$(4+n)$-dimensional spacetime. In the case of the scenario of Large Extra
Dimensions \cite{ADD}, in which the additional spacelike dimensions are compact
with size $L$, the latter feature introduces a new fundamental higher-dimensional
scale for gravity $M_*$, that can be significantly lower than the 4-dimensional scale $M_P$, thus
increasing the fundamental Newton's constant -- and  the strength of
gravitational interactions -- by orders of magnitude.

An immediate consequence of the above assumption is that particles with
energies larger than $M_*$ can probe physics beyond the scale of quantum gravity
while their collisions may trigger the appearance of strong gravitational
phenomena and lead to the creation of extended objects ($p$-branes, string balls,
string states, etc), rather than ordinary particles, including tiny black holes
\cite{creation}. This type of trans-planckian particle collision could be
achieved in various environments, such as at ground-based colliders \cite{colliders}
or in high energy cosmic-ray interactions in the atmosphere of the Earth \cite{cosmic}
(for an extensive discussion of the phenomenological implications and additional
references, see the reviews \cite{Kanti, reviews, Harris}). The black
holes produced with horizon radius $r_h \ll L$ are submerged into the higher-dimensional
spacetime, which modifies their properties (horizon radius, temperature, entropy,
life-time etc.) \cite{ADMR}. Nevertheless, they are still expected to go through the
same stages in their life as their 4-dimensional counterparts: a short {\it balding}
phase, during which the black hole will shed all additional quantum numbers apart
from its mass, angular momentum and charge, the more familiar {\it spin-down} phase,
during which the black hole will lose mainly its angular momentum, then the
{\it Schwarzschild} phase, when the black hole gradually loses its
actual mass, and finally the {\it Planck} phase, where the black hole has
reduced to a quantum object with unknown properties.

During the two intermediate phases, the spin-down and Schwarzschild, the black
hole loses energy through the emission of Hawking radiation \cite{hawking}
(and through superradiance during the former phase). This will take place via the
emission of elementary particles both in the bulk and on the brane, and it will
be characterised by a very distinct thermal spectrum. It is this emission
spectrum that upon successful detection will provide the most convincing
evidence not only for the creation of microscopic black holes but of the very
existence of extra spacelike dimensions. For this reason, the
emission spectra from a higher-dimensional black hole have been under
intensive study over the last few years. Having the simplest
gravitational line-element, the Schwarzschild phase was the first one
to undergo a systematic study both analytically \cite{kmr1, Frolov1, kmr2}
and numerically \cite{HK1}. Both bulk (scalars) and brane (scalars, fermions
and gauge bosons) channels were studied, and the derived emission spectra
were shown to depend strongly on the number of additional spacelike dimensions
in nature, both in terms of the amount of energy and number of particles produced
per unit time and of the type of particles produced. More general
spherically-symmetric backgrounds were also studied, such as
Schwarzschild-de Sitter \cite{BGK}, Schwarzschild-Gauss-Bonnet \cite{Barrau}
and Reissner-Nordstrom \cite{Jung-charge} backgrounds, as well as the effect
of the mass of the emitted particles on the spectrum \cite{mass}. Recently,
the graviton emission rates by a higher-dimensional Schwarzschild
black hole were also investigated by a number of authors \cite{gravitons}.

After the spherically-symmetric Schwarzschild phase was almost exhaustively
studied, the investigation of the form of the radiation spectra for a
higher-dimensional rotating black hole was initiated. Two early analytical
studies \cite{Frolov2,IOP1} were supplemented, during the last year or so,
by a number of works attempting to shed light on this more technically
involved case. Exact numerical results for the emission of Hawking radiation
from a $(4+n)$-dimensional rotating black hole, in the form of scalar fields
on the brane, were first presented in \cite{HK2}. In that work, the exact value of the
angular eigenvalue, that couples the angular and radial part of the equation
of motion of the scalar field and does not exist in closed form, was found
numerically, and the radial equation was also solved by employing numerical
analysis. A comprehensive study of the fluxes of particles, energy and
angular momentum for the emission of scalar fields on the brane soon
followed \cite{DHKW}, that once again presented exact results for the
emission rates including the angular distribution of particles. A similar
extensive analysis, focussed on the emission of gauge bosons on the brane by a
rotating black hole, appeared in \cite{CKW}. In both of the latter works,
it was shown that all emission rates for scalars and gauge fields are
greatly enhanced by both the dimension of spacetime and the angular
momentum of the black hole. Results produced independently in \cite{IOP2, IOP3},
for these two species of particles, reveal a good agreement between the two
sets of analyses. Additional studies in the literature featured the study of the
superradiance effect for a higher-dimensional black hole \cite{super}
and of the emission in the bulk \cite{rot-bulk}.

Recently, and while this work was in progress, a study \cite{IOP3}
produced power and angular momentum emission
spectra from a rotating black hole for the last species of brane-localised
particles, i.e. fermions. The radial part of the spinor equation was
solved numerically with the angular eigenvalue given by an analytical expansion
(with terms kept up to 6th order). The power and angular momentum emission spectra
were found to exhibit the same characteristics as the ones for scalars and
gauge bosons. The present work aims to complete the above analysis by
presenting all emission spectra (particle, energy and angular momentum)
as well as total emissivities (emission rates integrated over the frequency)
and angular distribution of the emitted radiation. To this end, both the
radial and angular part of the fermionic equation of motion will be solved
numerically, and an exact numerical value for the angular eigenvalue will
also be determined and employed in the calculation.

The outline of our paper is as follows: we start, in Section 2, with the
presentation of the theoretical framework for our analysis, and the derivation
of the radial and angular part of the spin-1/2 equation of motion on the brane.
The expressions for the transmission coefficients and the various Hawking
radiation emission rates, adopted for emission on the brane, are reviewed in
Section 3. In Section 4, the numerical techniques used to integrate the angular and
radial parts of the spin-1/2 equation of motion and determine the angular eigenvalue
are described. Section 5 presents our numerical results for
the transmission coefficient, the particle, energy and angular momentum fluxes,
as well as the total emissivities, and their dependence on both the dimension
of spacetime and angular momentum of the black hole is investigated. The angular
distribution of the particle and energy emission rates are also studied by
using the exact values of the spin-1/2-weighted spheroidal harmonics. We
finish with a summary of our results and conclusions, in Section 6.


\section{Fermion field in a brane-induced rotating black hole background}
\label{thf}

We start our analysis by presenting the line-element that describes the
4-dimensional gravitational background on a brane placed in the vicinity of a
higher-dimensional rotating, uncharged black hole. The latter $(4+n)$-dimensional
spacetime is given by the well-known Myers-Perry  solution \cite{MP}. The
induced-on-the-brane gravitational background follows by projecting the
higher-dimensional one on the brane, by fixing the values of all additional
azimuthal coordinates to $\pi/2$. Then, the brane line-element assumes the
form\footnote{In order for the black hole to be considered a classical object,
its mass $M_{BH}$ will be assumed to be at least a few times larger than
the fundamental scale of gravity $M_*$. Then, under the assumption that the
brane self-energy is of the order of $M_*$, its effect on the gravitational
background can be considered negligible. Analyses that address the complete
bulk-brane-black-hole system have only led, up to now, to consistent brane
trajectories solutions \cite{traj} -- for an attempt to analytically construct
a small black hole stuck on a tensional brane, see \cite{Nemanja}.} \cite{Kanti}
\begin{equation}
\begin{split}
ds^2=\left(1-\frac{\mu}{\Sigma\,r^{n-1}}\right)dt^2&+\frac{2 a\mu\sin^2\theta}
{\Sigma\,r^{n-1}}\,dt\,d\varphi-\frac{\Sigma}{\Delta}dr^2 \\[3mm] &\hspace*{-1cm}
-\Sigma\,d\theta^2-\left(r^2+a^2+\frac{a^2\mu\sin^2\theta}{\Sigma\,r^{n-1}}\right)
\sin^2\theta\,d\varphi^2\,,
\end{split} \label{induced}
\end{equation}
where
\begin{equation}
\Delta = r^2 + a^2 -\frac{\mu}{r^{n-1}}\,, \qquad
\Sigma=r^2 +a^2\,\cos^2\theta\,.
\label{master}
\end{equation}
The mass and angular momentum (transverse to the $r \varphi$-plane) of the black hole
are then given by
\begin{equation}
M_{BH}=\frac{(n+2) A_{n+2}}{16 \pi G}\,\mu\,,  \qquad
J=\frac{2}{n+2}\,M_{BH}\,a\,, \label{def}
\end{equation}
with $G$ being the $(4+n)$-dimensional Newton's constant, and $A_{n+2}$
the area of an $(n+2)$-dimensional unit sphere, given by
\begin{equation}
A_{n+2}=\frac{2 \pi^{(n+3)/2}}{\Gamma[(n+3)/2]}\,.
\end{equation}
The 4-dimensional background (\ref{induced}) is the one `felt' by all Standard
Model fields, including fermions, which are restricted to live on the brane.
As the reader notes, the higher-dimensional black hole has been assumed to
have only one angular momentum component, instead of the $[(n+3)/2]$ possible
ones. The reason for this is that, under the assumption that the brane is
infinitely-thin, the colliding particles have a non-vanishing impact parameter
only along the usual three spacelike coordinates; therefore, the resulting
angular momentum will be around an axis also in our brane. An upper bound can
be imposed on this sole angular momentum parameter of the black hole by
demanding the creation of the black hole from the collision of the two
particles. The maximum value of the impact parameter between the two particles
that can lead to the creation of a black hole was found to be \cite{Harris}
\begin{equation}
b_\text{max}=2 \,\biggl[1+\biggl(\frac{n+2}{2}\biggr)^2\biggr]^{-\frac{1}{(n+1)}}
\mu^{\frac{1}{(n+1)}}\,.
\end{equation}
This, in conjunction with the relation $J=b M_{BH}/2$ \cite{IOP1} for the
angular momentum of the black hole, leads to
\begin{equation}
a^\text{max}_*=\frac{n+2}{2}\,,
\label{amax}
\end{equation}
where $a_* \equiv a/r_{h}$. For $n \geq 1$, the black hole horizon is given by
the unique solution of the equation $\Delta(r)=0$, and may be conveniently
written as $r_{h}^{n+1}=\mu/(1+a_*^2)$.

Using the Newman-Penrose formalism, the Kinnersley tetrad and Boyer-Lindquist
co-ordinates, Teukolsky \cite{Teukolsky} decoupled the field equations for spin
$s=0,1/2,1$ and 2 field perturbations of a Petrov Type D (vacuum) background metric
(that also includes the traditional 4-dimensional Kerr spacetime). Teukolsky
expressed all the spin-field equations as one single `master' equation, with the
`helicity', or spin-weight, $h=(+|s|,-|s|)$ of the field as a parameter. As
Teukolsky's original equation does not apply to the case of propagation in a
projected-on-the-brane higher-dimensional background, as the one given
in Eq. (\ref{induced}), here we have re-derived this equation. The new
`brane master' equation now describes spin $s=0,1/2$ and 1 field perturbations
of both `helicities', and has the form
\begin{equation} \label{eq:Teuk.eq.}
\begin{aligned}
&\left[\frac{(r^2+a^2)^2}{\Delta}-a^2\sin^2\theta\right]\parddiff{\Omega_{h}}{t}+
\frac{2a\mu}{\Delta r^{n-1}}\pparddiff{\Omega_{h}}{t}{\phi}+
\left[\frac{a^2}{\Delta}-\frac{1}{\sin^2\theta}\right]\parddiff{\Omega_{h}}{\phi}\,- \\[1.5mm]
&\Delta^{-h}\frac{\partial}{\partial r}\left(\Delta^{h+1}\pardiff{\Omega_{h}}{r}\right)-
\frac{1}{\sin\theta}\frac{\partial}{\partial \theta}\left(\sin\theta\pardiff{\Omega_{h}}
{\theta}\right)-
2h\left[\frac{a\Delta'(r)}{2\Delta}+\frac{i\cos\theta}{\sin^2\theta}\right]
\pardiff{\Omega_{h}}{\phi}\,+ \\[1.5mm]
&2h\left[r+\bar\rho -\frac{(r^2+a^2)\Delta'(r)}{2\Delta}\right]\pardiff{\Omega_{h}}{t}+
h\left[h\cot^2\theta-1+(2-\Delta''(r))\delta_{h,|h|}\right]\Omega_{h}=
\Sigma\cdot T_{h}\,,
\end{aligned}
\end{equation}
where $\bar\rho=r+i a\cos\theta$.
In the above, $\Omega_{h}=\Omega_{h}(t,r,\theta,\phi)$ represents the spin-field
perturbation and $T_{h}$ the source term in the field equations - we refer the reader
to \cite{Kanti, Teukolsky} for details on the derivation of the above equation, and
for the precise definitions of $\Omega_{h}(t,r,\theta,\phi)$ and
$T_{h}$, in each spin case.
Equation (\ref{eq:Teuk.eq.}) is formally the same as Teukolsky's original
four-dimensional equation, except for the different expression of the metric
function $\Delta(r)$ which is now $n$-dependent, and the term with $\Delta''(r)$
which is identically zero in four dimensions (this was already noted in~\cite{CKW}
for the spin-1 case) and, obviously, for $h<0$.

The `master' equation (\ref{eq:Teuk.eq.}) is clearly separable for any value of
the `helicity' $h$, so that its solution can be
written as a sum over the Fourier modes:
\begin{equation} \label{eq:Fourier expansion for Omega_h}
\begin{aligned}
\Omega_{h}(t,r,\theta,\phi)&=\int_{-\infty}^{+\infty}d\omega\sum_{l=|h|}^{+\infty}
\sum_{m=-l}^{+l}{}_{h}a_{\Lambda}\ {}_{h}\Omega_{\Lambda}(t,r,\theta,\phi)  \\
{}_{h}\Omega_{\Lambda}(t,r,\theta,\phi)&={}_{h}R_{\Lambda}(r){}_{h}S_{\Lambda}(\theta)
e^{-i\omega t}e^{+im\phi}\,,
\end{aligned}
\end{equation}
where ${}_{h}a_{\Lambda}$ are the Fourier coefficients, and the set of `quantum'
numbers is denoted by $\Lambda\equiv \{\ell m\omega\}$.

The radial and angular ODEs resulting from the separation of variables of the
`master' equation are, respectively:
\begin{equation} \label{eq:radial teuk. eq.}
\Delta^{-h}\frac{d}{dr}\left(\Delta^{h+1}\frac{d{}_{h}R_{\Lambda}}{dr}\right)+
\left[\frac{K^2-ihK\Delta'(r)}{\Delta}+4ih \omega r+h(\Delta''(r)-2)
\delta_{h,|h|}-{}_{h}\lambda_{\Lambda}\right]{}_{h}R_{\Lambda}=0
\end{equation}
where we have used the definition
\begin{equation}
K = (r^2 + a^2) \omega - a m,
\end{equation}
and
\begin{eqnarray}
\label{eq:ang. teuk. eq.}
&~& \hspace*{-2.5cm} \left[
\frac{d}{dx}\left((1-x^2)\frac{d}{dx}\right)+a^2\omega^2(x^2-1)-2h a\omega x \,-
\right. \nonumber \\[1mm] & & \hspace*{2cm} \left.
\frac{(m+hx)^2}{1-x^2}+{}_{h}\lambda_{\Lambda}+2ma\omega+h\right]{}_{h}S_{\Lambda}(x)
=  0\,,
\end{eqnarray}
where $x\equiv \cos \theta$. In the above, ${}_{h}\lambda_{\Lambda}$ is the constant
of separation between the angular and radial equations.

In the spin-1/2 case, the wave function of the fermion is represented by two
spinors $P^A$ and $\bar Q^{A'}$ \cite{Chandrasekhar}, that, in the massless case,
satisfy the equations $\nabla_{AA'} P^A=\nabla_{AA'} Q^A=0$. By using the
Kinnersley tetrad $(\ell_\mu, n_\mu, m_\mu,  m_\mu ^*)$ and the Fourier
decomposition (\ref{eq:Fourier expansion for Omega_h}), these equations
can be separated into first-order ODEs with respect to the $\theta$ and $r$
coordinates. For the 2-component spinor $P^A=(P^0, P^1)$,
we make the decomposition
\begin{align}
P^0 &= (r - ia\cos \theta)^{-1} {}_{-1/2}R_{\Lambda}(r)\,\,{}_{-1/2}S_{\Lambda}(\theta)\,e^{im\phi} e^{-i\omega t}
\\[1mm]
P^1 &= {}_{+1/2}R_{\Lambda}(r)\,\,{}_{+1/2}S_{\Lambda}(\theta)\,e^{im\phi} e^{-i\omega t}.
\end{align}
Defining new radial functions
\begin{equation}
\Pplus  = \Delta^{1/2}\,{}_{+1/2}R_{\Lambda}, \qquad
\Pminus = {}_{-1/2}R_{\Lambda}\,,
\end{equation}
the first-order radial equations take the form \cite{Chandrasekhar}
\begin{align} \label{coupled radial}
\Delta^{1/2} \left( \partial_r - i K / \Delta \right) \Pminus &=  \lambda \Pplus ,\nonumber \\
\Delta^{1/2} \left( \partial_r + i K / \Delta \right) \Pplus &=  \lambda \Pminus ,
\end{align}
where $\lambda = \sqrt{{}_{-1/2}\lambda_{\Lambda}}$. It may easily be shown
that similar definitions for the components of the $\bar Q^{A'}$ spinor lead
to the same system of first-order radial ODEs. The coupled radial equations
(\ref{coupled radial}) may then be written in matrix form as follows
\begin{equation}
\frac{d \bP}{dr} = \left( \frac{i K}{\Delta} \bsig_3 + \frac{\lambda}
{\Delta^{1/2}} \bsig_1 \right) \bP\,,
\label{rad-eq-matrix}
\end{equation}
where $\bP = \left( \begin{smallmatrix} \Pminus \\ \Pplus \end{smallmatrix} \right)$,
and $\bsig_1 = \left( \begin{smallmatrix} 0 & 1 \\ 1 & 0 \end{smallmatrix} \right)$
and $\bsig_3 = \left( \begin{smallmatrix} 1 & 0 \\ 0 & -1 \end{smallmatrix} \right)$
are the Pauli spin matrices. By taking the complex conjugate of this equation it is
clear that, if $\bP$ is a solution, then so is $\bsig_1 \bP^\ast$. This transformation
corresponds to time-reversal. On causal grounds we will only admit one set of
solutions: those that are ingoing at the horizon. This alternative, but
equivalent to Eq. (\ref{eq:radial teuk. eq.}), set of radial equations
will be used for the numerical determination of the radial part of the
spinor wave function, as we will see in detail in Section 4.

\section{Hawking radiation in the form of fermion fields}
\label{Hawking}

We are interested in the Hawking radiation of an evaporating black
hole on the brane, which is described by the (``past'') Unruh vacuum
$\left| U^{-} \right\rangle $.
The quantum field theory of fermions on the Kerr black hole has been
extensively studied in the literature \cite{Leahy,Unruh1,Vilenkin,Bolash}, but for
ease of reference, we now outline the key features.
For more details of the general formalism relating to fermions on curved space-time,
see \cite{Brill}, or \cite{Groves} for a more recent review.

In constructing the state $\left| U^{-} \right\rangle $, it is most straightforward
to use Dirac 4-spinors, in which case the wavefunction of the massless fermionic field (that,
for historical reasons, we call `neutrino') takes the form \cite{Leahy,Unruh1,Vilenkin}
\begin{equation}
\psi = \frac {e^{-i\omega t} e^{im \varphi }}{ {\cal {F}}}
\left(
\begin{array}{c}
\eta \\ L \eta
\end{array}
\right) ,
\label{psimode}
\end{equation}
where
\begin{equation}
{\cal {F}} = \left[ \Delta \sin ^{2} \theta\ \bar \rho ^{2}
\right] ^{\frac {1}{4}}.
\end{equation}
Here $L$ is the lepton number \cite{Vilenkin}, so that $L=+1$ for neutrinos and $L=-1$ for
anti-neutrinos.
The two-spinor $\eta $ takes the form \cite{Leahy,Unruh1,Vilenkin}
\begin{equation}
\eta = \left(
\begin{array}{c}
R_{1} (r) S_{1}(\theta) \\
R_{2} (r) S_{2} (\theta )
\end{array}
\right) .
\end{equation}
The angular functions $S_{1}$ and $S_{2}$ satisfy the equations \cite{Leahy,Unruh1,Vilenkin}
\begin{eqnarray}
\left[ \frac {d}{d\theta } + \left( a\omega \sin \theta - \frac {m}{\sin \theta } \right) \right]
S_{1} (\theta ) & = & \lambda S_{2}(\theta ) ,\\
\left[ \frac {d}{d\theta } - \left( a \omega \sin \theta - \frac {m}{\sin \theta } \right) \right]
S_{2} (\theta ) & = & -\lambda S_{1} (\theta ) ,
\end{eqnarray}
and are related to the spin-weighted spheroidal harmonics ${}_{h}S_{\Lambda }$
(\ref{eq:ang. teuk. eq.}) by \cite{Leahy}
\begin{equation}
S_{1} (\theta )   =   \left( \frac {\sin \theta }{2} \right) ^{\frac {1}{2}}
{}_{-\frac {1}{2}}S_{\Lambda }(\cos \theta ) ,
\qquad
S_{2} (\theta )  =  \left( \frac {\sin \theta }{2} \right) ^{\frac {1}{2}}
{}_{\frac {1}{2}}S_{\Lambda }(\cos \theta ) .
\end{equation}
There is a subtlety in the definition of the radial functions $R_{1}$, $R_{2}$,
depending on whether we are considering neutrinos or anti-neutrinos.
For neutrinos ($L=+1$), we have \cite{Leahy,Unruh1,Vilenkin}
\begin{equation}
R_{1} (r) = P_{-},
\qquad
R_{2}(r) = P_{+},
\label{neutrinomodes}
\end{equation}
but for anti-neutrinos ($L=-1$), the radial functions swap over \cite{Vilenkin}:
\begin{equation}
R_{1} (r) = P_{+},
\qquad
R_{2} (r) = P_{-}.
\label{antineutrinomodes}
\end{equation}
This is particularly important when calculating the particle fluxes for neutrinos and anti-neutrinos.

The usual ``in'' and ``up'' modes are defined by the behaviour of the radial functions
near the horizon and at infinity \cite{Leahy,Unruh1}:
\begin{eqnarray}
\left( P_{-}^{\rm {in}}, P_{+}^{\rm {in}} \right)
& = &
\frac{1}{\sqrt{8 \pi^2}}
\left\{
\begin{array}{lcl}
\left( A^{\rm {in}}_{\Lambda } e^{i\omega r_{*}},
e^{-i\omega r_{*}} \right) & & r\rightarrow \infty ,
\\[0.5mm]
\left( 0, B^{\rm {in}}_{\Lambda } e^{-i{\tilde {\omega }} r_{*}} \right)
& & r\rightarrow r_{h} ,
\end{array}
\right.
\label{inmodes}
\\[1mm]
\left( P_{-}^{\rm {up}}, P_{+}^{\rm {up}} \right)
& = &
\frac{1}{\sqrt{8 \pi^2}}
\left\{
\begin{array}{lcl}
\left( B^{\rm {up}}_{\Lambda } e^{i\omega r_{*}} , 0 \right)
& & r\rightarrow \infty ,
\\[0.5mm]
\left( e^{i{\tilde {\omega }} r_{*}} ,
A^{\rm {up}}_{\Lambda } e^{-i{\tilde {\omega }} r_{*}} \right)
& & r \rightarrow r_{h} ,
\end{array}
\right.
\label{upmodes}
\end{eqnarray}
and correspond to modes originating either at past infinity or past horizon,
respectively. In the above, $r_{*}$ is the standard ``tortoise'' co-ordinate, defined by
\begin{equation} \label{tortoise}
\frac {dr_{*}}{dr} = \frac {r^{2}+a^{2}}{\Delta (r)},
\end{equation}
and
\begin{equation}
{\tilde {\omega }} = \omega - m \Omega ,
\end{equation}
with $\Omega $ the horizon angular velocity
\begin{equation}
\Omega = \frac {a_{*}}{\left( 1 + a_{*}^{2} \right) r_{h}} .
\end{equation}
In Eqs. (\ref{inmodes})-(\ref{upmodes}),
$A^{\rm {in}}_{\Lambda }$, $B^{\rm {in}}_{\Lambda }$,
$A^{\rm {up}}_{\Lambda }$, $B^{\rm {up}}_{\Lambda }$ are constants
of integration which satisfy the Wronskian relations \cite{Unruh1}
\begin{equation}
1- \left| A_{\Lambda }^{\bullet } \right| ^{2} =
\left| B_{\Lambda }^{\bullet }  \right| ^{2},
\qquad
B^{\rm {up}^*}_{\Lambda } = - B^{\rm {in}}_{\Lambda } ,
\label{Wronskianrelations}
\end{equation}
where $\bullet $ is either ``in'' or ``up''.
These relations are useful for simplifying the flux formulae we shall derive in this section.

The (``past'') Unruh vacuum $\left| U^{-} \right\rangle $ is defined in the usual way, by
taking the ``in'' modes to have positive frequency with respect to co-ordinate time near infinity
(which corresponds to $\omega >0$) and using a set of modes having positive frequency with
respect to Kruskal time in the vicinity of the past event horizon.
Using the lemma in Appendix H of \cite{Novikov}, a suitable set of modes
having positive frequency with respect to Kruskal time near the past horizon is
\begin{equation}
\left[ 2 \cosh \left( \frac {{\tilde {\omega }}}{2T_{\text{H}}} \right) \right] ^{-\frac {1}{2}}
\left\{ \exp \left( \frac {{\tilde {\omega }}}{4T_{\text{H}}} \right) \psi ^{\rm {up}}_{\Lambda }
+ \exp \left( \frac {-{\tilde {\omega }}}{4T_{\text{H}}} \right) \psi  ^{\rm {down}^*}_{\Lambda }
\right\}\,,
\end{equation}
for {\em {any}} value of ${\tilde {\omega }}$.
The modes $\psi  ^{\rm {down}}_{\Lambda }$ are given by the complex conjugate of the modes
$\psi  ^{\rm {up}}_{\Lambda }$, after changing the sign of the Kruskal co-ordinates.
Similarly, a suitable set of modes
having {\em negative} frequency with respect to Kruskal time near the past horizon is
\begin{equation}
\left[ 2 \cosh \left( \frac {{\tilde {\omega }}}{2T_{\text{H}}} \right) \right] ^{-\frac {1}{2}}
\left\{ \exp \left( \frac {{-\tilde {\omega }}}{4T_{\text{H}}} \right) \psi ^{\rm {up}}_{\Lambda }
+ \exp \left( \frac {{\tilde {\omega }}}{4T_{\text{H}}} \right) \psi  ^{\rm {down}^*}_{\Lambda }
\right\}\,,
\end{equation}
for any value of ${\tilde {\omega }}$.
The ``down'' modes are vanishing on the right-hand quadrant of the conformal diagram of the
space-time, which is the region in which we are interested.
Therefore we do not consider them further.
Further details of this construction can be found in \cite{Leahy,Novikov,Unruh2}.
We therefore find that a suitable
expansion for the neutrino field in terms of our positive and negative frequency modes is (cf. \cite{Leahy}):
\begin{eqnarray}
\hspace*{-2.5cm}\Psi & = &
\sum _{\ell =1/2}^{\infty } \sum _{m=-\ell }^{\ell } \Bigg\{
\int _{\omega >0} d\omega \left[
\psi ^{\rm {in}}_{{\Lambda }} {{\hat {a}}}^{{\rm {in }}}_{\Lambda }
+ \psi ^{\rm {in}^*}_{{-\Lambda }} {{\hat {b}}}^{{\rm {in }}\dag  }_{\Lambda }
\right]
\nonumber \\ & &
+ \int_{{\rm {all}} \, {\tilde {\omega }}} d{\tilde {\omega }}
\left[ 2 \cosh \left( \frac {{\tilde {\omega }}}{2T_{H}} \right)  \right] ^{-\frac {1}{2}}
\left[
\exp \left( \frac {{\tilde {\omega }}}{4T_{H}}\right) {\hat {a}}^{\rm {up}}_{\Lambda }
+
\exp \left( \frac {-{\tilde {\omega }}}{4T_{H}}\right) {\hat {b}}^{\rm {up} \dag }_{\Lambda }
\right]\psi ^{\rm {up}}_{\Lambda }
\Bigg\} ,
\label{Psiexpansion}
\end{eqnarray}
where $-\Lambda\equiv \{\ell,-m,-\omega\}$ and $T_\text{H}$ is the Hawking temperature
of the $(4+n)$-dimensional, rotating black hole, given by
\begin{equation}
T_\text{H}=\frac{(n+1)+(n-1)\,a_*^2}{4\pi\,(1+a_*^2)\,r_{h}}\,.
\label{temp}
\end{equation}
The Fourier coefficients ${}_ha_{\Lambda}$ in (\ref{eq:Fourier expansion for Omega_h}) have been promoted to quantum
operators ${\hat {a}}_{\Lambda }$ (where we have dropped the helicity subindex $h$ for clarity).
The state $\left| U^{-} \right\rangle $ is then annihilated by the operators
${\hat {a}}^{\bullet }_{\Lambda }$ and ${\hat {b}}^{\bullet }_{\Lambda }$:
\begin{equation}
{\hat {a}}^{\rm {in}}_{\Lambda } \left| U^{-} \right\rangle
= {\hat {a}}^{\rm {up}}_{\Lambda } \left| U^{-} \right\rangle =0
= {\hat {b}}^{\rm {in}}_{\Lambda } \left| U^{-} \right\rangle
= {\hat {b}}^{\rm {up}}_{\Lambda } \left| U^{-} \right\rangle.
\end{equation}
We now have the machinery in place to compute the fluxes of particles, energy and angular momentum
from the black hole.

Firstly, the particle flux is given by \cite{Leahy,Unruh1}
\begin{equation}
\frac {dN}{dt}  =
\int _{S_{\infty }} d\theta \, d\varphi \, r^{2} \sin \theta \,
\langle U^{-} | J^{r} | U^{-} \rangle \, ,
\end{equation}
where $J^{r}$ is the radial component of the conserved current \cite{Unruh1}:
\begin{equation}
J^{\mu } = \frac{1}{2}\left[{\bar {\Psi }},\gamma ^{\mu } \Psi \right].
\end{equation}
Here, the Dirac adjoint ${\bar {\Psi }}$ is defined by \cite{Unruh1,Brill}:
\begin{equation}
{\bar {\Psi }} = \Psi ^{\dag } \gamma ^{0} ,
\end{equation}
where $\gamma ^{0}$ is the flat-space Dirac matrix.
A suitable representation of the curved-space Dirac matrices $\gamma ^{\mu }$ for
the space-time metric (\ref{induced}) can be found in \cite{Unruh1}.
We compute the fluxes for neutrinos $\nu$ and antineutrinos $\bar {\nu}$ , using the field expansion (\ref{Psiexpansion})
and the mode functions of the form (\ref{psimode}).
This gives the total particle flux per unit time to be \cite{Leahy}:
\begin{equation}
\frac {d^{2}N_{\nu + {\bar {\nu }}}}{d(\cos \theta ) dt}
=
\frac{1}{4\pi}
\sum _{\ell = 1/2}^{\infty } \sum _{m=-\ell }^{\ell } \int_{\omega = 0}^{\infty }
\frac {d\omega }{\exp\left(\tilde{\omega}/T_\text{H}\right) + 1}
{\mathbb {T}}_{\Lambda }
 \left({}_{-1/2}S_{\Lambda}^2+{}_{+1/2}S_{\Lambda}^2\right) .
\label{totalparticle}
\end{equation}
The transmission coefficient ${\mathbb {T}}_{\Lambda } $ is the transmitted proportion of the flux which,
in the case of an ``up'' mode (\ref{upmodes}) is equal to
\begin{equation}
{\mathbb {T}}_{\Lambda } = \left| B^{\rm {up}}_{\Lambda } \right| ^{2} .
\end{equation}
By virtue of the Wronskian relations (\ref{Wronskianrelations}) this is the same as the absorption
probability of an ``in'' mode.
Similarly, the net particle flux is \cite{Leahy}
\begin{equation}
\frac {d^{2}N_{\nu - {\bar {\nu }}}}{d(\cos \theta ) dt}
=
\frac{1}{4\pi}
\sum _{\ell = 1/2}^{\infty } \sum _{m=-\ell }^{\ell } \int_{\omega = 0}^{\infty }
\frac {d\omega }{\exp\left(\tilde{\omega}/T_\text{H}\right) + 1}
{\mathbb {T}}_{\Lambda }
 \left({}_{-1/2}S_{\Lambda}^2-{}_{+1/2}S_{\Lambda}^2\right) .
 \label{netparticle}
\end{equation}
It is tempting, in view of Eqs. (\ref{totalparticle}) and (\ref{netparticle}) to view the contribution from
neutrinos as being proportional to ${}_{-1/2}S_{\Lambda}$ and that from anti-neutrinos as being
proportional to ${}_{+1/2}S_{\Lambda}$.
In fact, at infinity, this is the case, but only because of the form of the ``in'' and ``up'' modes in this
limit and Eqs. (\ref{neutrinomodes})-(\ref{antineutrinomodes}).
It is not true more generally.

The fluxes of energy and angular momentum per unit time are given in terms of components of the
stress-energy tensor:
\begin{eqnarray}
\frac {dE}{dt} & = &
\int _{S_{\infty }} d\theta \, d\varphi \, r^{2} \sin \theta \,
\langle U^{-} | T^{tr} | U^{-} \rangle ,
 \\
\frac {dJ}{dt} & = &
\int _{S_{\infty }} d\theta \, d\varphi \, r^{2} \sin \theta \,
\langle U^{-} | T^{r}_{\varphi } | U^{-} \rangle\, ,
\end{eqnarray}
where the stress-energy tensor of the neutrino field is given by \cite{Unruh1}
\begin{equation}
T_{\mu \nu } = \frac {i}{4} \Big(
\left[{\bar {\Psi }}, \gamma _{\left( \mu \right. } \nabla _{\left. \nu \right) } \Psi\right]
- \left[ \nabla _{\left( \mu \right. } {\bar {\Psi }}, \gamma _{\left. \nu \right) }\Psi\right]
\Big) .
\end{equation}
The spinor covariant derivative is defined by \cite{Unruh1}
\begin{equation}
\nabla _{\mu } \Psi = \left( \partial _{\mu } - \Gamma _{\mu } \right) \Psi ,
\end{equation}
where the $\Gamma _{\mu }$ matrices can be taken to be traceless and satisfy \cite{Unruh1}
\begin{equation}
\gamma ^{\mu }{}_{;\nu } - \Gamma _{\nu } \gamma ^{\mu } + \gamma ^{\mu } \Gamma _{\nu } = 0.
\end{equation}
These can be straightforwardly computed using the formula (cf. \cite{Cardall},
but with different conventions)
\begin{equation}
\Gamma _{\mu } = -\frac {1}{8} g_{\nu \sigma } \left[
\gamma ^{\sigma }, \gamma ^{\nu }{}_{;\mu } \right] .
\end{equation}
The formulae for the matrices $\Gamma _{\mu }$ are sufficiently large that we
do not reproduce them here.

After computing the expectation value of the stress-energy tensor in the state
$\left| U^{-} \right\rangle $, the total fluxes of particles, energy and angular
momentum emitted by the black hole per unit time and frequency, in the form of
fermion modes, are given by \cite{Leahy}
\begin{eqnarray}
\frac {d^{2}N_{\nu+\bar \nu}}{dt  d\omega}  & = &
\frac {1}{2\pi} \sum _{\ell =1/2}^{\infty }
\sum _{m=-\ell }^{\ell }
\frac {1}{\exp\left(\tilde{\omega}/T_\text{H}\right) + 1}
{\mathbb {T}}_{\Lambda }\,,
\label{flux} \\
\frac {d^{2}E}{dt  d\omega }  &  = &
\frac {1}{2\pi} \sum _{\ell =1/2}^{\infty }
\sum _{m=-\ell }^{\ell }
\frac {\omega }{\exp\left(\tilde{\omega}/T_\text{H}\right) + 1}
{\mathbb {T}}_{\Lambda }\,,
\label{power} \\
\frac {d^{2}J}{dt  d\omega}  & = &
\frac {1}{2\pi} \sum _{\ell =1/2}^{\infty }
\sum _{m=-\ell }^{\ell }
\frac {m}{\exp\left(\tilde{\omega}/T_\text{H}\right) + 1}
{\mathbb {T}}_{\Lambda }\,.
\label{ang-mom}
\end{eqnarray}
The emission rates (\ref{flux})-(\ref{ang-mom}) describe the different
fluxes emitted by the black hole over the whole solid angle $\Omega_2^2=4 \pi$,
and follow by performing an integration over the angular coordinates
$\theta$ and $\phi$. If we go one step backwards, we may derive the
angular distribution of the emitted radiation by displaying the exact
dependence of the differential rates on the azimuthal angle $\theta$,
and write
\begin{eqnarray}
\frac {d^{3}N_{\nu+\bar{\nu}}}{d(\cos\theta)dt  d\omega}  & = &
\frac {1}{4\pi} \sum _{\ell =1/2}^{\infty }
\sum _{m=-\ell }^{\ell }
\frac {1}{\exp\left(\tilde{\omega}/T_\text{H}\right) + 1}
{\mathbb {T}}_{\Lambda }
 \left({}_{-1/2}S_{\Lambda}^2+{}_{+1/2}S_{\Lambda}^2\right)\,,
\label{fluxang} \\
\frac {d^{3}E}{d(\cos\theta)dt  d\omega }  &  = &
\frac {1}{4\pi} \sum _{\ell =1/2}^{\infty }
\sum _{m=-\ell }^{\ell }
\frac {\omega }{\exp\left(\tilde{\omega}/T_\text{H}\right) + 1}
{\mathbb {T}}_{\Lambda }
\left({}_{-1/2}S_{\Lambda}^2+{}_{+1/2}S_{\Lambda}^2\right)\,.
\label{powerang}
\end{eqnarray}
It is easy to check that the above two expressions reduce to Eqs.
(\ref{flux})-(\ref{power}) when we use the angular function normalization
\begin{equation}
\int_{-1}^{1} \left|{}_{\pm1/2} S_\Lambda (\cos \theta)  \right|^2 d(\cos \theta)
= \int_{-1}^{1} \left| {}_{\pm1/2} S_\Lambda(x) \right|^2 d x  = 1\,.
\end{equation}
Note that in the particle (\ref{fluxang}) and power (\ref{powerang}) fluxes, both
`helicities' $h=-1/2$ and $h=+1/2$ make a contribution. As we will see, this will
lead to radiation spectra that are invariant under the parity transformation
$\theta \to \pi-\theta$ (for more discussions on the asymmetry under parity
of the derived spectra in purely four dimensions, see, e.g.
\cite{Leahy,Vilenkin,Bolash}).



\section{Numerical analysis}
\label{numerical}

The transmitted flux in each angular mode may be calculated by
numerically solving the radial equations. Before we may begin,
however, we must first solve the angular equations to determine the
eigenvalue $\lambda = \sqrt{{}_{-1/2}\lambda_\Lambda}$ that appears in
the radial equations. In order to compute the angular distributions of
the particle and energy fluxes, we will also need the angular
eigenvectors: the spin-1/2-weighted spheroidal harmonics. In this section,
we briefly outline the numerical methods used to solve both the
angular and radial equations, and describe our method for determining
the transmission coefficient $\mathbb{T}_{\Lambda}$.

The angular eigenvalues ${}_{h}\lambda_{\Lambda}$ were calculated using a
spectral decomposition method, described in a paper by Hughes \cite{Hughes-2000}.
In appendix A of \cite{Hughes-2000}, it is shown that spheroidal harmonics may
be expanded as a sum over spherical harmonics of the same spin-weight,
\begin{equation}
{}_hS_{\Lambda}(\theta) = \sum^\infty_{j=l_{\rm{min}}} {}_h b_{lj}^{a \omega}
{}_hY_{jm}( \theta )\,,
\label{spectral-decomp}
\end{equation}
where $l_{\min} = \text{max}(|h|, |m|)$. A recurrence relation for the expansion coefficients
${}_h b_{lj}^{a\omega}$ is generated by substituting (\ref{spectral-decomp})
into the angular equation (\ref{eq:ang. teuk. eq.}) and exploiting the
orthogonality properties of the spin-weighted spherical harmonics. Writing
$b_j \equiv {}_h b_{lj}^{a\omega}$ for clarity, the result is
\begin{eqnarray}
 - \, {}_h \mathcal{E}_{jm\omega} \, b_j  & = &
\left[ (a\omega)^2 c^m_{j-2,j,2} \right] b_{j-2} + \left[ (a\omega)^2 c^m_{j-1,j,2} -
2 a \omega h c^m_{j-1,j,1} \right] b_{j-1} \, \nn \\ & &
+ \left[ (a \omega)^2 c^m_{j,j,2} - 2 a \omega h c^m_{j,j,1} - j(j+1) \right] b_j
\nn \\ & &
+ \left[ (a \omega)^2 c^m_{j+1,j,2} - 2 a \omega h c^m_{j+1,j,1} \right] b_{j+1} \,
+\left[ (a \omega)^2 c^m_{j+2,j,2} \right] b_{j+2} ,
\,
\label{hughes-recurrence}
\end{eqnarray}
where we have defined
\begin{align}
c^m_{j,l,2} &= \frac{2}{3} \sqrt{\frac{2j+1}{2l+1}} \langle j,2,m,0 | l,m \rangle
\langle j,2,-h,0 | l,-h \rangle + \frac{1}{3} \delta_{jl}  , \label{CM-inner-c2} \\[1mm]
c^m_{j,l,1} &= \, \, \, \sqrt{\frac{2j+1}{2l+1}} \langle j,1,m,0 | l,m \rangle
\langle j,1,-h, 0 | l, -h \rangle \label{CM-inner-c1} .
\end{align}
Here, $\langle j,i,m,0 | l,n \rangle$ are Clebsch-Gordan coefficients. The
separation constant ${}_h \mathcal{E}_{\Lambda}$, appearing on the left-hand-side of
Eq. (\ref{hughes-recurrence}), is related to the angular parameter
${}_{h}\lambda_{\Lambda}$ by
\begin{equation}
{}_h \mathcal{E}_{\Lambda} = {}_{h}\lambda_{\Lambda} + 2ma \omega - a^2 \omega^2 + h(h+1).
\end{equation}
In the spin-half case, $\ell$ and $m$ are half-integers, and $h = -1/2$. In the
non-rotating limit the solution reduces to $\mathcal{E}_{\ell m} = \ell(\ell+1)$ and
${}_{-1/2}\lambda_{\Lambda} = (\ell+\tfrac{1}{2})^2$.

The recurrence relations (\ref{hughes-recurrence}) may be rewritten in matrix form.
The eigenvalue of the matrix is (minus) the separation constant ${}_h \mathcal{E}_{\Lambda}$,
and the eigenvector is a column of expansion coefficients $b_j$. Since the matrix
is band-diagonal, the eigenproblem may be solved in a numerically efficient way.
For more details on this method, see Appendix A in \cite{Hughes-2000}.

The results for the angular eigenvalues were independently checked using a partial
fraction method \cite{Leaver-1985}, and a variant of the `shooting' method~\cite{Num.Rec.},
which is described in detail in~\cite{CKW,Casals:2004zq}. The angular eigenvalues
coincided up to double precision. The values of the corresponding separation constants
${}_{h}\mathcal{E}_{\Lambda}$ following from the above analysis are
shown in Fig. \ref{spheroidal}(a), for some indicative spin-1/2 modes.

\FIGURE{
\mbox{
\includegraphics[height=5cm,clip]{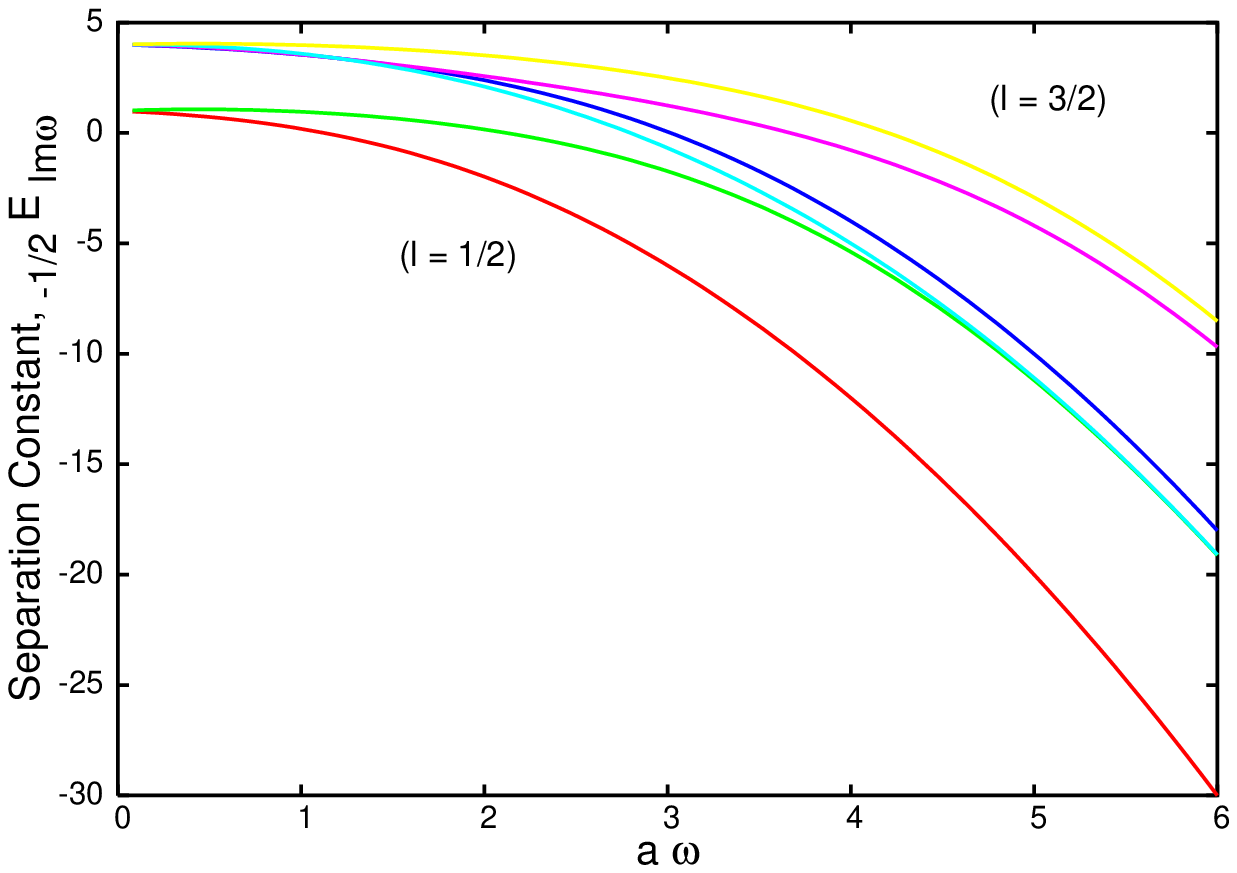}}
{\includegraphics[height=5cm,clip]{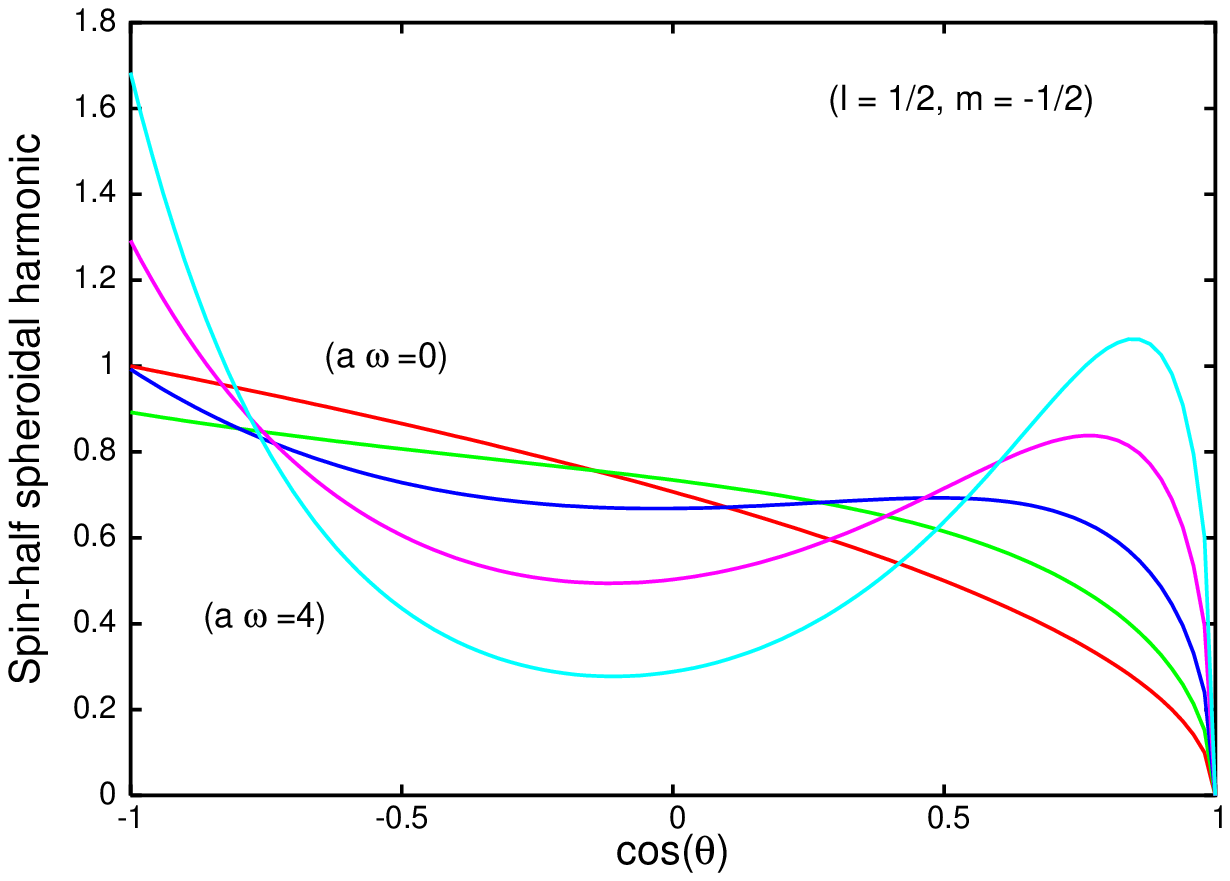}}
\caption{{\bf (a)} The separation constant ${}_{h}\mathcal{E}_{\Lambda}$,
for the modes $\ell=1/2,3/2$ and $m=-\ell, ..., \ell$, and `helicity' $h=-1/2$; {\bf (b)}
Spin-1/2 weighted spheroidal harmonics for a spinor field with $h=-1/2$, $\ell=1/2$,
$m=-1/2$, and for $a \omega=(0,1,2,3,4)$. }
\label{spheroidal}
}

The spectral decomposition method may also be used to calculate the angular
functions ${}_{h}S_{\Lambda}(x)$, according to Eq. (\ref{spectral-decomp}).
However, this calculation would first require the determination of the spin-weighted
spherical harmonics ${}_hY_{jm}( \theta )$. Instead, we preferred to follow the
series expansion approach introduced by Leaver \cite{Leaver-1985}. A solution to
Eq. (\ref{eq:ang. teuk. eq.}) may be expressed as
\begin{equation}
{}_{h}S_{\Lambda}(x) = e^{a\omega x} (1+x)^{\tfrac{1}{2} |m - h|}
(1-x)^{\tfrac{1}{2} |m+h|} \sum_{n=0}^\infty c_n (1+x)^n,
\end{equation}
where $x = \cos(\theta)$ and the expansion coefficients $c_n$ are related by a
three-term recurrence relation,
\begin{align}
\alpha_0 c_1 + \beta_0 c_0 &= 0, \\[1mm]
\alpha_n c_{n+1} + \beta_n c_n + \gamma_n c_{n-1} &= 0, \quad n = 1, 2, \ldots
\end{align}
The recurrence coefficients $\alpha_n, \beta_n, \gamma_n$ are
\begin{align}
\alpha_n &= -2 (n + 1) (n + 2k_- + 1) \nn \\[1mm]
\beta_n  &= n(n-1) + 2n(k_- + k_+ + 1 - 2 a\omega) \nn \\
          & \quad -\left[2 a \omega (2 k_- + h + 1) - (k_- + k_+)(k_- + k_+ + 1) \right]
      -\left[a^2 \omega^2 + {}_h\mathcal{E}_{\Lambda} \right] \nn \\[1mm]
\gamma_n &= 2 a \omega (n + k_- + k_+ + h) ,
\end{align}
with $k_- = \tfrac{1}{2}|m-h|$ and $k_+ = \tfrac{1}{2}|m + h|$.
This expansion, together with the eigenvalue ${}_h \mathcal{E}_{\Lambda}$ calculated
via the spectral decomposition, provides an efficient method for the calculation
of spin-half spheroidal harmonics, such as those shown in Fig.~\ref{spheroidal}(b).
Note that ${}_{-1/2}S^2_{\Lambda}$ is clearly greater on the Southern hemisphere
than on the Northern one, with a similar (but opposite) asymmetry arising also
for ${}_{+1/2}S^2_{\Lambda}$. However, their sum is invariant under the change
$\theta \to \pi-\theta$, which means that, unless both
`helicities' $h=\pm 1/2$ are taken into account in the calculation of the
angular distribution of the radiation fluxes (\ref{fluxang}) and (\ref{powerang}),
the resulting fermionic emission will be asymmetric with respect to parity.

Once the eigenvalues of the angular equation are known, we may proceed
to solve the radial equations. One possible approach is to start from
the second-order radial Teukolsky equation (\ref{eq:radial
teuk. eq.}), and introduce a new radial ``tortoise'' coordinate (like the one in (\ref{tortoise})) to
transform to a potential barrier problem
\begin{equation}
\left[\frac{d^2}{{dr_* }^2} - V(r)\right] R(r) = 0.
\end{equation}
Equations of this form are familiar from quantum mechanical problems,
and standard methods for finding transmission coefficients may be
employed. However, a disadvantage of this method is that the
definition of the tortoise coordinate depends on the number of extra
dimensions, and the relation $r_*(r)$ can be hard to invert (see the discussion in \cite{CKW}). Here,
we prefer instead to work with the pair of coupled first-order
equations (\ref{rad-eq-matrix}). These may be solved by combining
series expansions at the horizon and at infinity with numerical
integration, as described below.

Briefly, our method for determining the transmission coefficient $\mathbb{T}_{\Lambda}$
is as follows. We start by demanding that the solution be purely ingoing at the horizon.
The ingoing solution may be written as a power series in $(r/r_h - 1)^{1/2}$ (see next
subsection). We evaluate this series solution at tenth order at a point close to the
horizon ($r/r_h - 1 \le 10^{-4}$) and use this as the initial condition for a numerical
integration scheme. We integrate outwards into the asymptotically-flat region,
$r/r_h \ge 100$, using a fourth-order Runge-Kutta routine. Here, the solution is
matched onto ingoing and outgoing power series solutions in $r_h/r$ which are also
evaluated at tenth order. The asymptotic solution at infinity is a linear sum of
ingoing and outgoing waves,
\begin{equation}
\bP \rightarrow \Ain\,\bP^\text{in} + \Aout\,\bP^\text{out}, \quad \text{ as \,}
r \rightarrow +\infty ,
\end{equation}
where $\Ain$ and $\Aout$ are complex constants, and  $\bP^\text{in}$ and
$\bP^\text{out} = \bsig_1 \bP^{\text{in}\ast}$ are series expansions described
below. Using our numerical solution we may determine $\Ain$ and $\Aout$ up to
an overall magnitude and phase.
The transmission coefficient is then simply
\begin{equation}
\mathbb{T}_{\Lambda}
= 1 - \left| \frac{\Aout}{\Ain} \right|^2.
\end{equation}

\subsection{Expansion at the horizon}

We may write $K$ and $\Delta$ as power series around $r = r_h$,
\begin{align}
K &= \sum_{k = 0}^2 K_k \left(\frac{r}{r_h} - 1\right)^k , & \Delta &= \sum_{k = 1}^\infty \Delta_k \left(\frac{r}{r_h}-1\right)^k  .
\end{align}
The first terms are
\begin{align}
K_0      &= (r_h^2+a^2) \omega - a m , &  K_1     &= 2 \omega r_h^2,\\[1mm]
\Delta_1 &= 2r_h^2 + (r_h^2+a^2)(n-1)  ,      & \Delta_2 &= r_h^2 - \tfrac{1}{2} n (n-1)(r_h^2+a^2) .
\end{align}
Defining $\eta \equiv (r/r_h - 1)^{1/2}$, the radial equations (\ref{rad-eq-matrix})
may be written as
\begin{equation}
\frac{d \bP}{d\eta}  =
\frac{1}{\eta}\,\bM(\eta)\,\bP ,
\end{equation}
where
\begin{equation}
\bM(\eta) = \sum_{k = 0}^\infty \bM_k \,\eta^k =
\left( 2is_0 \right) \bsig_3 + \eta \left( \tfrac{2 \lambda r_h}{\sqrt{\Delta_1}} \right)
\bsig_1 + \eta^2 2is_0 \left( \tfrac{K_1}{K_0} - \tfrac{\Delta_2}{\Delta_1} \right) \bsig_3
+ \ldots
\end{equation}
and $s_0 = K_0 r_h / \Delta_1$. This form makes it clear that $r = r_h$ is a regular singular point of the radial equations.

The ingoing solution at the horizon may then be written as
\begin{equation}
\bP^\text{in} =
\begin{bmatrix} \Pminus \\[1mm]  \Pplus  \end{bmatrix} =
\left(\frac{r}{r_h} - 1\right)^{-i s_0} \left[  \begin{array}{l}
 \frac{2 \lambda r_h}{ \sqrt{ \Delta_1 } (1 - 4is_0)}\,\eta  +  \ldots  \\[2mm]
 1 + \left( \frac{2 \lambda^2 r_h^2}{\Delta_1 (1 - 4is_0)} - i s_0 \left( \tfrac{K_1}{K_0} - \tfrac{\Delta_2}{\Delta_1} \right) \right)  \,\eta^2
+ \ldots
\end{array} \right] .
\label{ingoing-soln-hor}
\end{equation}
Note that the $\Pminus$ series expansion includes only odd powers of $\eta$, whereas
the $\Pplus$ expansion includes only even powers. The outgoing solution may be
generated by time-reversal, that is, $\bP^\text{out} = \bsig_1 \bP^{\text{in}\ast}$.

\subsection{Expansion at infinity}
Similar techniques may be applied to find an expansion around the singular point at
infinity. First, we define $z \equiv r_h/r$. The radial equations can then be
expressed as
\begin{equation}
\frac{d \bP}{d z} = \frac{1}{z^2} \left( \sum_{k = 0}^\infty \bN_k \,z^k \right) \bP .
\end{equation}
This makes it clear that the point at infinity is an irregular singular point (of rank 1)
and we should look for a series expansion of the form
\begin{equation}
\bP = e^{t_0 z^{-1}} z^{- t_1} \sum_{k = 0}^\infty \evec_k \,z^k ,
\label{P-infty}
\end{equation}
where $t_0$ and $t_1$ are constants, and $\evec_k$ are two-component eigenvectors to
be determined. Substituting this into the radial equations and equating the
coefficients of like powers of $z$ yields a system of equations
\begin{align}
( \bN_0 + t_0 ) \evec_0 & =    0 \\
( \bN_0 + t_0 ) \evec_1 & =  - (\bN_1 + t_1) \evec_0 \\
( \bN_0 + t_0 ) \evec_2 & =  - (\bN_1 + (t_1 - 1)) \evec_1 - \bN_2 \evec_0 \\
\ldots &  & \nn \\
( \bN_0 + t_0 ) \evec_k & =  - (\bN_1 + (t_1 - (k-1)) \evec_{k-1} - \sum_{j=0}^{k-2}{\bN_{k-j} \evec_j} .
\label{infinity_eqns}
\end{align}
The solution for $t_0$ and $t_1$ depends on the number of extra dimensions ($n$).
In 4D, we
find
\begin{equation}
t_0 = \pm i \omega r_h, \quad t_1 = \pm i \omega (r_h^2 + a^2)/r_h ,
\end{equation}
with the positive choice of sign corresponding to an outgoing solution. In 4D
such a series solution is given by
\begin{eqnarray}
\bP^\text{out} & = & \begin{pmatrix} \Pminus \\ \Pplus \end{pmatrix}
\nonumber \\ & = &
e^{i \omega r} r^{i \omega (r_h^2 + a^2)/r_h }
\left( \begin{bmatrix} 1 \\ 0 \end{bmatrix} +
       \frac{i}{2 \omega r} \begin{bmatrix} \lambda^2 - 2 \omega^2 (r_h^2+a^2)^2 / r_h^2 + 2a\omega m \\ -\lambda \end{bmatrix}
       + \mathcal{O}\left( \frac{r_h^2}{r^2} \right)
\right) .
\end{eqnarray}
In the asymptotic limit, the upper component dominates over the lower
component in the outgoing solution. Again, the ingoing solution can
be found by interchanging $\Pminus$ and $\Pplus$ and taking the complex
conjugate.

In higher dimensions the index $t_0$ is unchanged, but $t_1 = 0$. The absence
of a logarithmic phase factor for higher-dimensional solutions indicates that the
influence of the gravitational potential is now ``short-range'' for large $r$
(up to $r \sim L$, where $L$ is the size of the extra dimensions). The potential
is still ``long-range'' near the horizon, however, and numerical troubles were avoided
by using series expansions at the horizon (and infinity) up to tenth order.


\section{Numerical results}
\label{num-results}

In this section, we present exact numerical results for the transmission coefficient
and the corresponding particle, energy and angular momentum emission rates for a
higher-dimensional, rotating black hole emitting spinor fields on the brane.
Their dependence on the dimension of spacetime $n+4$ and the angular momentum
parameter $a_*$ will be investigated, and the angular distribution of the emitted
number of particles and energy will also be derived. In the final part of this
section, the total emissivities of flux, power and angular momentum will be
computed, and their dependence on the fundamental parameters $n$ and $a_*$
will also be discussed.

\subsection{Transmission coefficient}

The first issue we need to address is the behaviour of the transmission
coefficient $\mathbb{T}_{\Lambda}$ in terms of both the angular
momentum parameter $a_{*}$ and the number of extra dimensions $n$. The
value of $\mathbb{T}_{\Lambda}$ follows from the numerical
integration of the radial equations (\ref{rad-eq-matrix}) after the exact
value of the eigenvalue $_{h}\lambda_{\Lambda}$ has been determined,
as described in the previous section.

\FIGURE{
\includegraphics[width=15cm,clip]{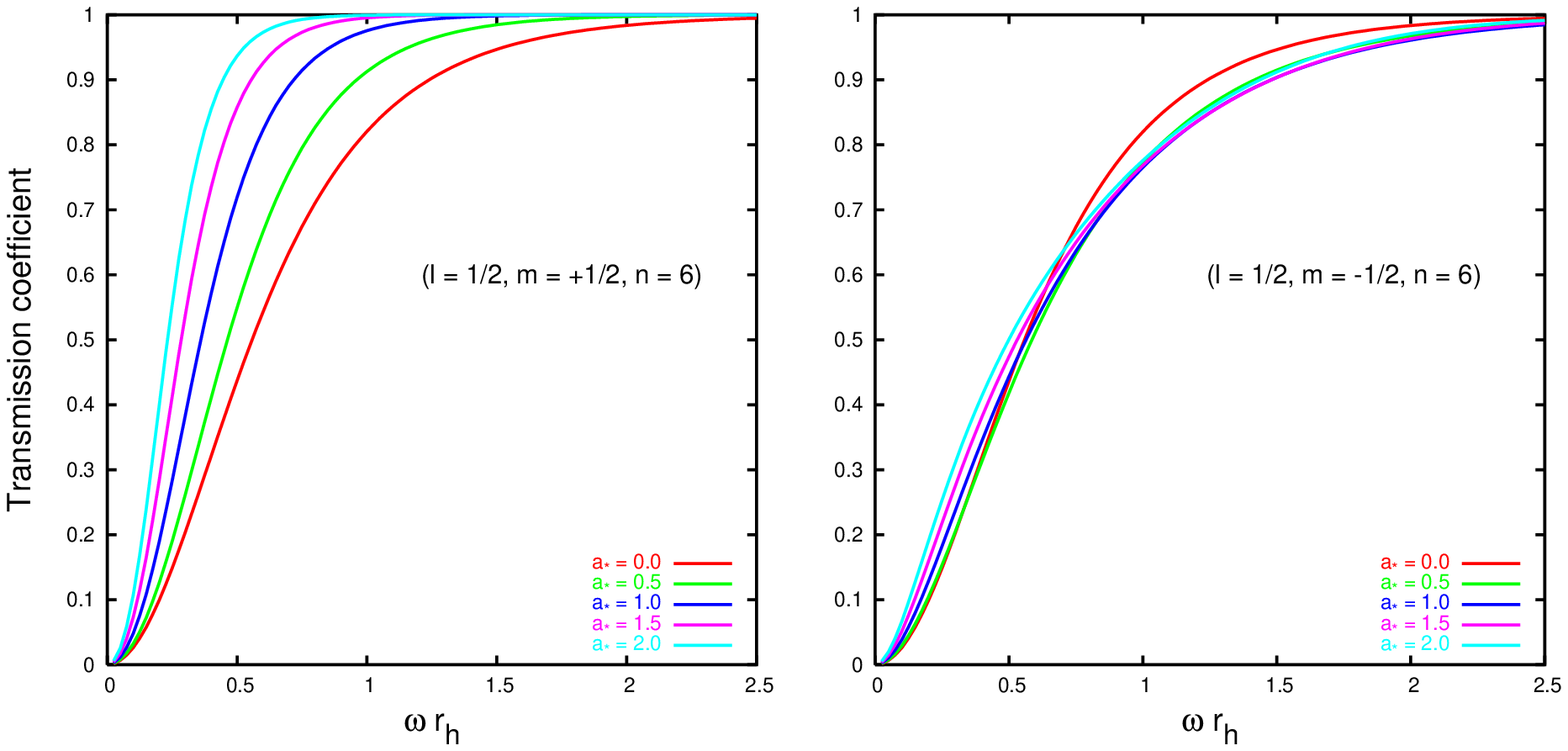}
\caption{Transmission coefficients for spin-1/2 emission on the
brane from a 10-dimensional black hole with variable $a_*$, for
the modes {\bf {(a)}} $\ell=1/2, m=1/2$, and {\bf {(b)}} $\ell=1/2, m=-1/2$.}
\label{transm-a} }
\FIGURE{
\thicklines
\includegraphics[width=7cm,height=6cm]{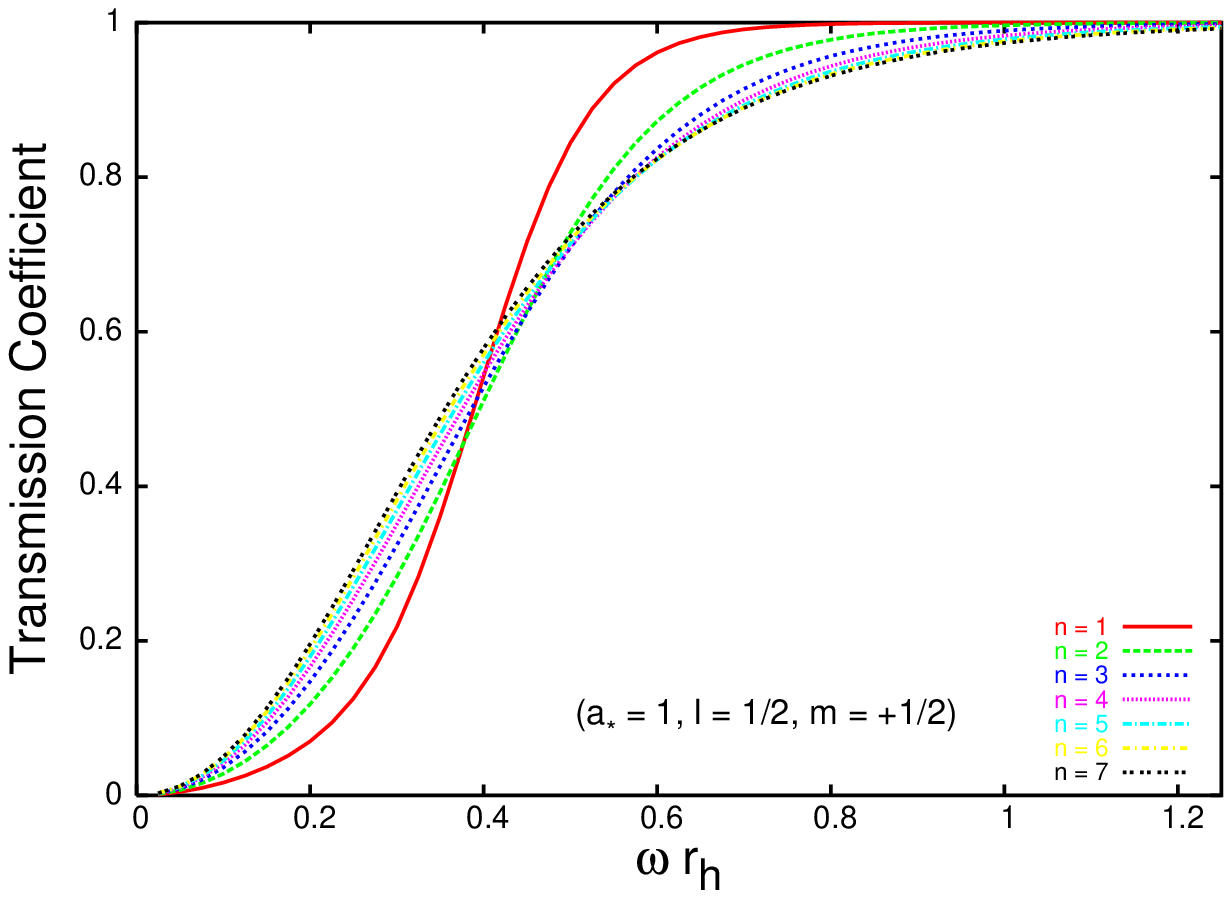} 
\includegraphics[width=7cm,height=6cm]{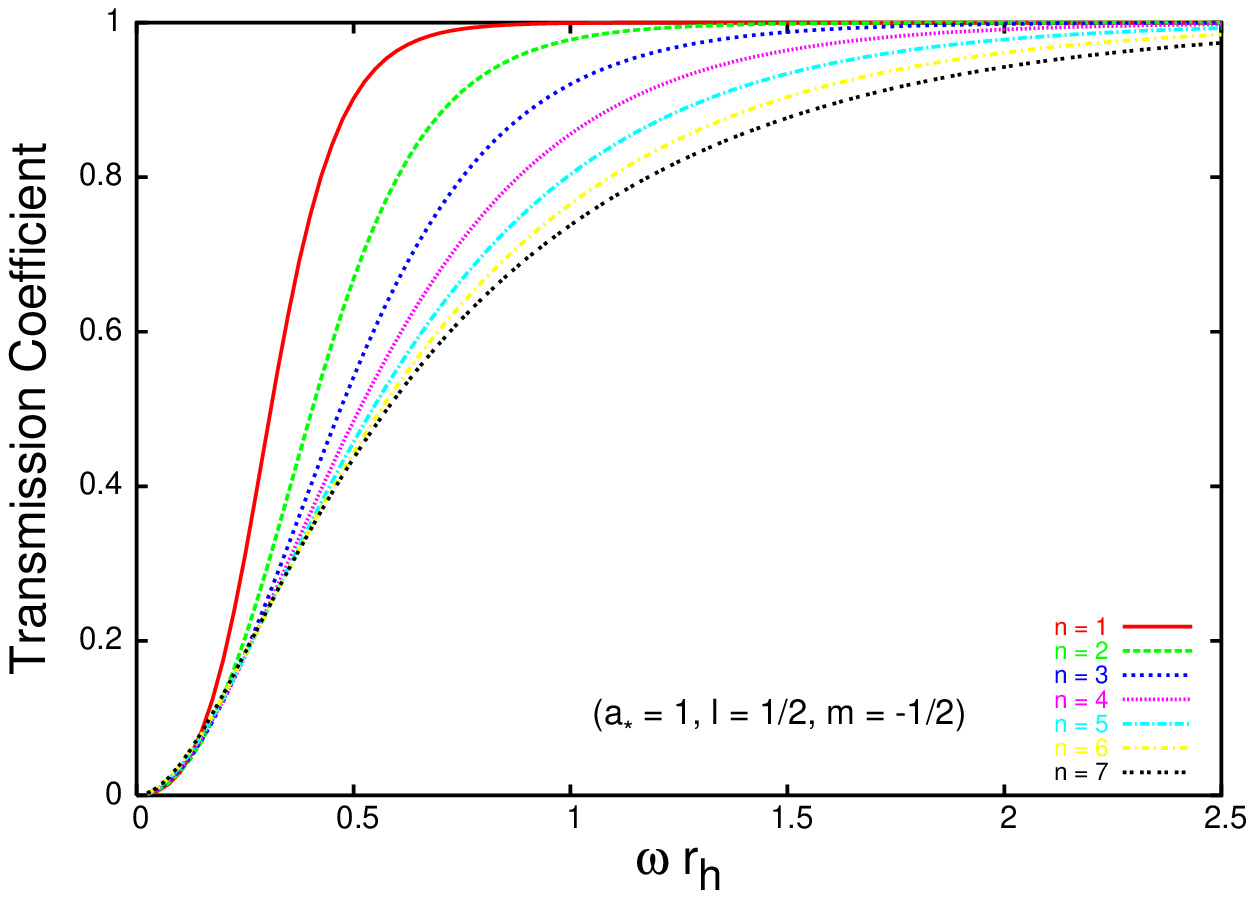}
\caption{Transmission coefficients for spin-1/2 emission on the brane from a black
hole with $a_*=1$ and variable $n$, for the modes {\bf {(a)}} $\ell=1/2, m=1/2$, and {\bf {(b)}}
$\ell=1/2, m=-1/2$.}
\label{transm-n}
}

The behaviour of the transmission coefficient also strongly depends  on
the particular mode studied, therefore in Figs. \ref{transm-a} and
\ref{transm-n} we present the dependence of  two indicative
modes ($\ell=1/2, m=1/2$) and ($\ell=1/2, m=-1/2$) on $a_*$ and $n$,
respectively.
Figure \ref{transm-a}(a) reveals the monotonic behaviour
of $\mathbb{T}_{\Lambda}$ for modes with $m>0$ that leads to a
continuous enhancement in terms of the angular momentum parameter $a_*$.
On the other hand, as we may see in Fig. \ref{transm-a}(b), modes
with $m<0$ exhibit a non-monotonic behaviour that strongly depends on
the energy regime that is probed: whereas the transmission coefficient
is enhanced in the low-energy regime, it is suppressed in the high-energy
one. The two panels of Fig. \ref{transm-n} depict the dependence of
$\mathbb{T}_{\Lambda}$ on the number of the transverse-to-the-brane
dimensions $n$. The exhibited behaviour is again found to be mode- and
energy-dependent. Modes with $m<0$ have a transmission coefficient
which is monotonically suppressed with $n$, while the behaviour of
modes with $m>0$ is characterised by enhancement in the low-energy
regime and suppression in the high-energy one.

A similar behaviour in terms of $a_*$ and $n$ was found for the corresponding
modes of the emitted gauge bosons studied in \cite{CKW}. The dependence of
modes with $m<0$ is easily observed to be identical in the two cases. For
modes with $m>0$, though, the agreement is not obvious since, in the case
of gauge bosons, modes with $m>0$ are superradiant ones. Nevertheless,
when the fact that a `more negative' value of $\mathbb{T}_{\Lambda}$
in the superradiant case is equivalent to a `more positive' value in
the non-superradiant one, agreement is again found between the two cases.
In \cite{CKW}, a detailed analysis was performed to show that the
above behaviour of the transmission coefficient was directly linked to
the behaviour of the effective potential seen by the propagating emitted
mode - as the behaviour found for fermions is similar to those for gauge
bosons, we will not repeat the same discussion here but, instead, refer the
reader to Ref. \cite{CKW} for more information on this point.


The spectrum of the emitted fermionic modes will also be strongly determined
by the behaviour of the thermal factor ($e^{\tilde\omega/T_\text{H}}+1)^{-1}$
in terms of the same parameters $n$ and $a_*$. It may be easily seen that, for
fixed $a_*$ and $r_h$, the temperature of the black hole increases with the
number of extra dimensions $n$, thus giving a boost to all emission spectra.
The dependence on the angular momentum parameter $a_*$, on the other hand, is much
more subtle, and depends on the particular mode and energy regime. For modes
with $m<0$, a simple numerical analysis reveals that the exponent
$\tilde \omega/T_\text{H}$ is an increasing function of $a_*$, for all
values of $\omega$, as long as $n \leq 1$, thus leading to a possible
suppression of the emission rates for these modes as $a_*$ increases;
however, for $n>1$, this exponent is found to be a decreasing function of
$a_*$, for low enough values of the energy and high enough values of the
angular momentum of the black hole. As the transmission coefficient
is also enhanced in these regimes, according to Fig. \ref{transm-a}(b),
fermionic modes with negative $m$ are more likely to be found in the low-energy
part of the spectrum of rapidly rotating black holes. Finally, for modes
with $m>0$, the exponent is found to be a decreasing function of $a_*$
in the low-angular-momentum regime, and an increasing function otherwise,
independently of the energy of the mode; as the transmission coefficient
is also monotonically enhanced with $a_*$, we expect these modes to mainly
dominate the radiation spectra of slowly rotating black holes.

\subsection{Flux emission spectra}

In the next three subsections, we present the various emission spectra for
spin-1/2 fields by using the definitions (\ref{flux})-(\ref{powerang})
and the exact numerical results for the transmission coefficient discussed
above. We start with the particle flux, i.e. the number of fermions emitted
by the black hole on the brane per unit time and unit frequency. Figures
\ref{n1-6flux} and \ref{a1flux} depict our results as a function of the
dimensionless energy parameter $\omega r_h$ and for fixed dimension
of spacetime and angular momentum of the black hole, respectively.

\FIGURE{
\thicklines
\includegraphics[height=5cm,clip]{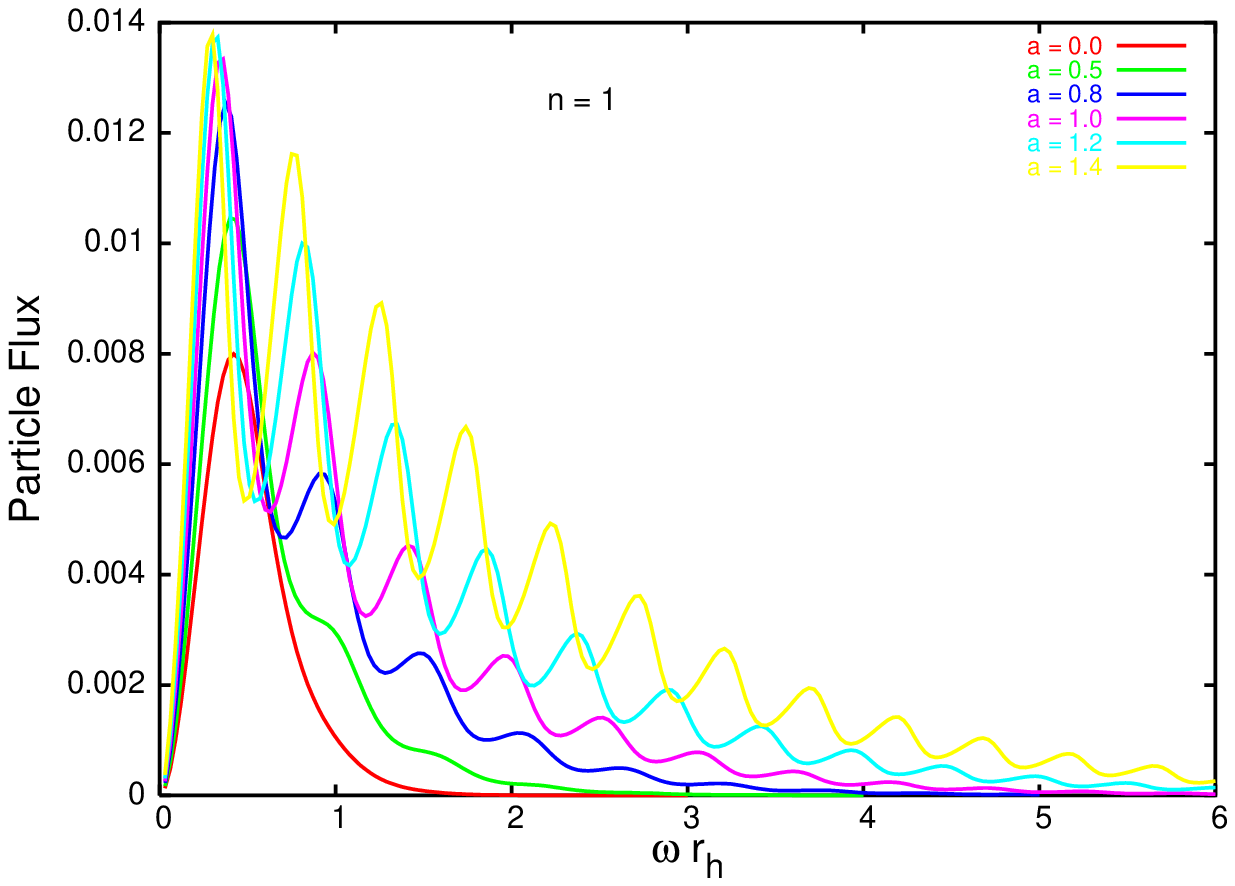}
\includegraphics[height=5cm,clip]{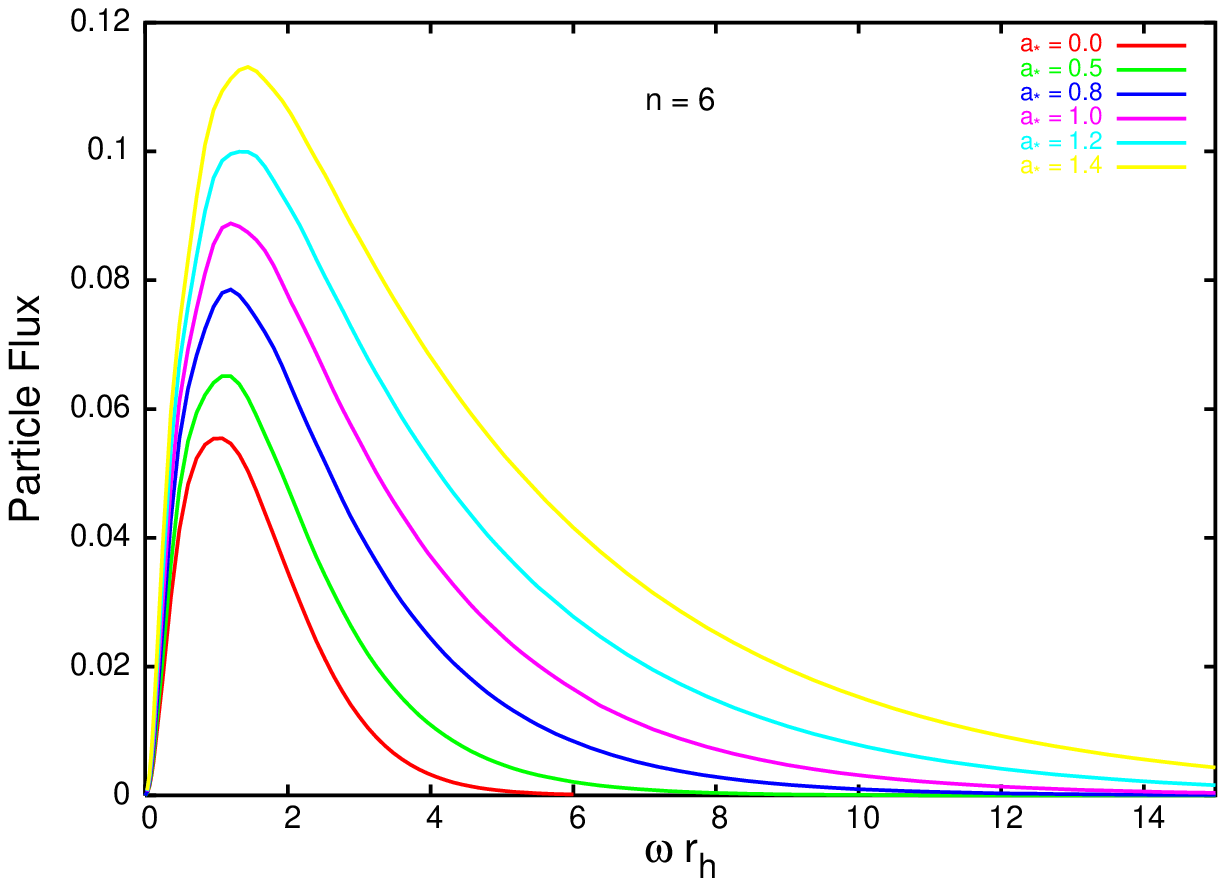}
\caption{Flux emission spectra for spin-1/2 particles on the brane from a rotating black hole,
for {\bf {(a)}} $n=1$, and {\bf {(b)}} $n=6$, and various values of $a_*$.\hspace*{1.5cm}}
\label{n1-6flux}
}
\FIGURE{
\parbox{12cm}{\begin{center}
\psfig{file=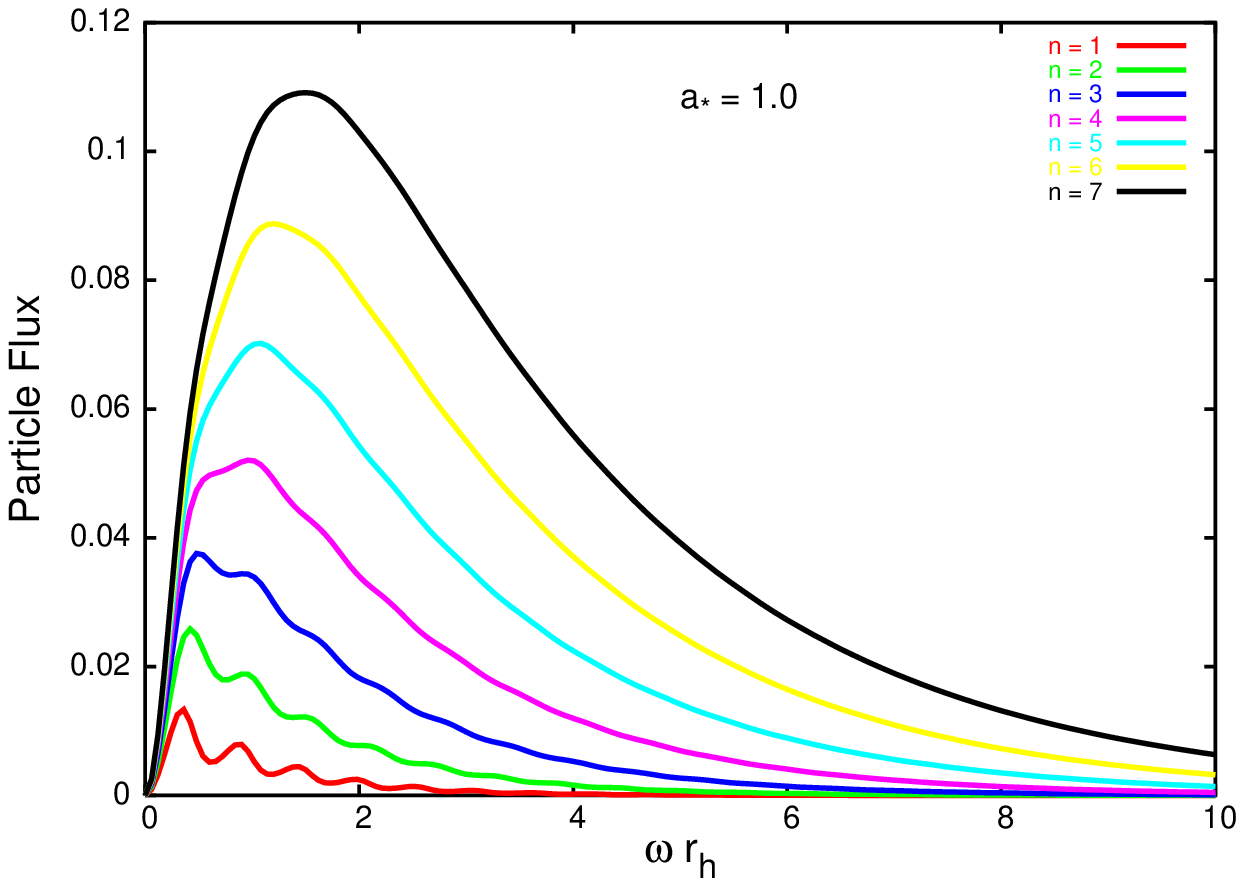, height=6cm}
\caption{Flux emission spectra for spin-1/2 particles on the brane from a rotating black hole,
with $a_*=1$, for various values of $n$.\hspace*{1.5cm}}
\label{a1flux}
\end{center}
}}
%

The dependence of the particle flux spectrum for fermions follows the same
pattern as the one for either scalars \cite{IOP1,DHKW} or gauge bosons \cite{CKW}.
As either the angular momentum of the black hole or the dimension of
spacetime increases, the number of particles emitted by the black hole is
greatly enhanced. According to Fig. \ref{n1-6flux}(a), for low values of $n$,
the black hole clearly tends to emit more low-energy particles than
high-energy ones, although as $a_*$ increases the number of the latter
becomes significant. For higher values of $n$, however, as Figs. \ref{n1-6flux}(b)
and \ref{a1flux} reveal, the black hole consistently prefers to emit particles
with energies $\omega r_h \geq 1$, with any increase of $a_*$ resulting
in the shift of the emission curve towards even larger values of $\omega r_h$.

The angular distribution of the flux spectra of fermionic modes -- characteristic
in the case of a rotating decaying black hole -- was also studied. It was found
to borrow features from the corresponding angular distributions of both scalars
and gauge bosons. The two indicative cases of $n=1$ and $n=6$, for three
different values of the angular momentum parameter, i.e. $a_*=(0, 0.6, 1)$, are
presented in Figs. \ref{n1f-ang} and \ref{n6f-ang}.
The depicted spectra correctly
reproduce the dependence on $a_*$ and $n$ discussed above, and in addition reveal
the existence of two forces that shape the final angular distribution pattern.
The centrifugal force, acting on all particles independently of their spin,
forces the emitted modes to concentrate on the equatorial plane, i.e. around
the value $\theta=\pi/2$ of the azimuthal angle. On the other hand, the
non-vanishing spin of the particle leads to a spin-rotation interaction with
the angular momentum of the black hole and tends to polarize radiation
along the rotation axis ($\theta=0,\pi$). The latter effect was, as expected,
absent in the
case of scalar fields \cite{DHKW} but quite prominent in the case of gauge
bosons \cite{CKW}. Here, as expected, it is also present but of a lesser
magnitude due to the smaller value of the spin.
According to Figs. \ref{n1f-ang}
and \ref{n6f-ang}, and in agreement with the behaviour found for gauge bosons,
the spin-rotation coupling is found to be particularly effective for particles
with low energy, but its effect becomes subdominant compared to the one of the
centrifugal force as either $a_*$ or $n$ increases -- the dying out of the
polarization of the radiation along the rotation axis, as the angular momentum
of the black hole increases, was also noted in the pure 4-dimensional case \cite{Leahy}.
\FIGURE{
\begin{tabular}{c} \hspace*{-0.4cm}
\epsfig{file=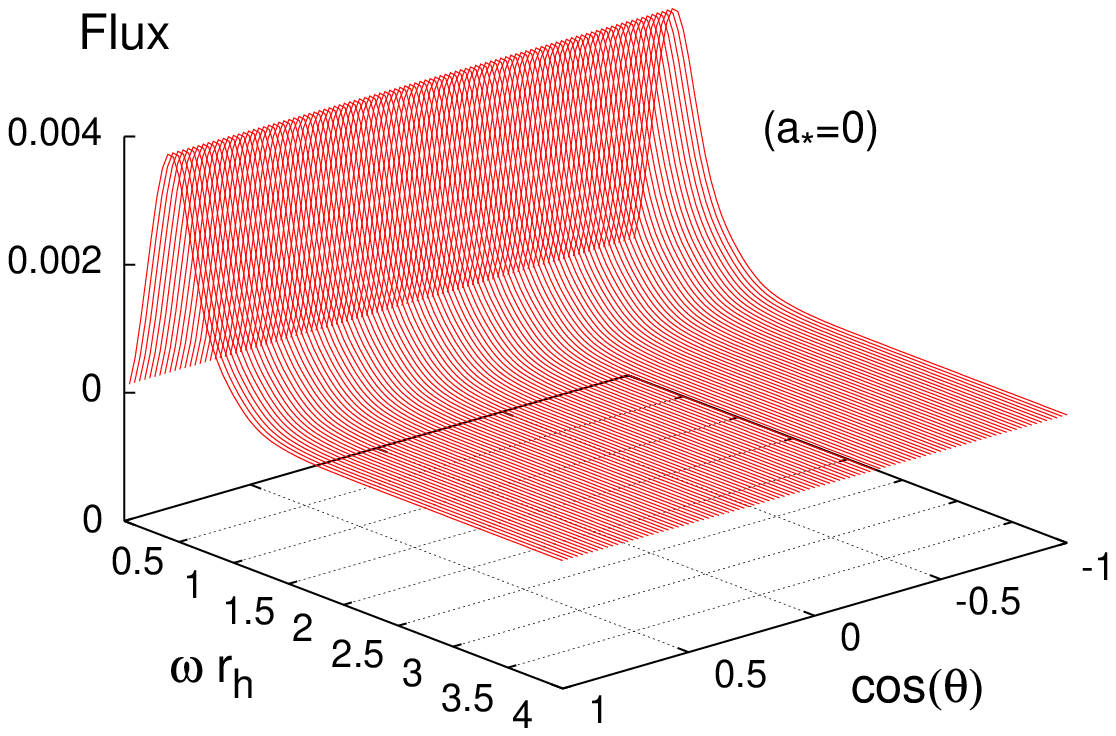, height=4.17cm}\hspace*{-1.15cm}
\epsfig{file=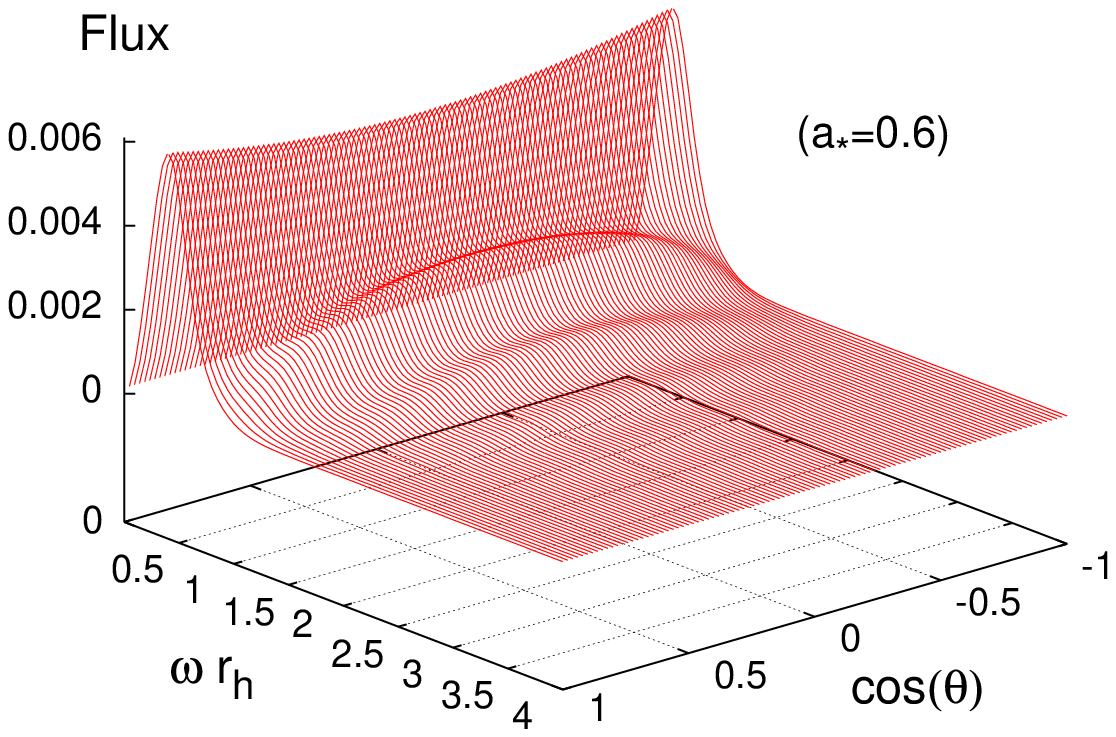, height=4.17cm}\hspace*{-1.15cm}{
\epsfig{file=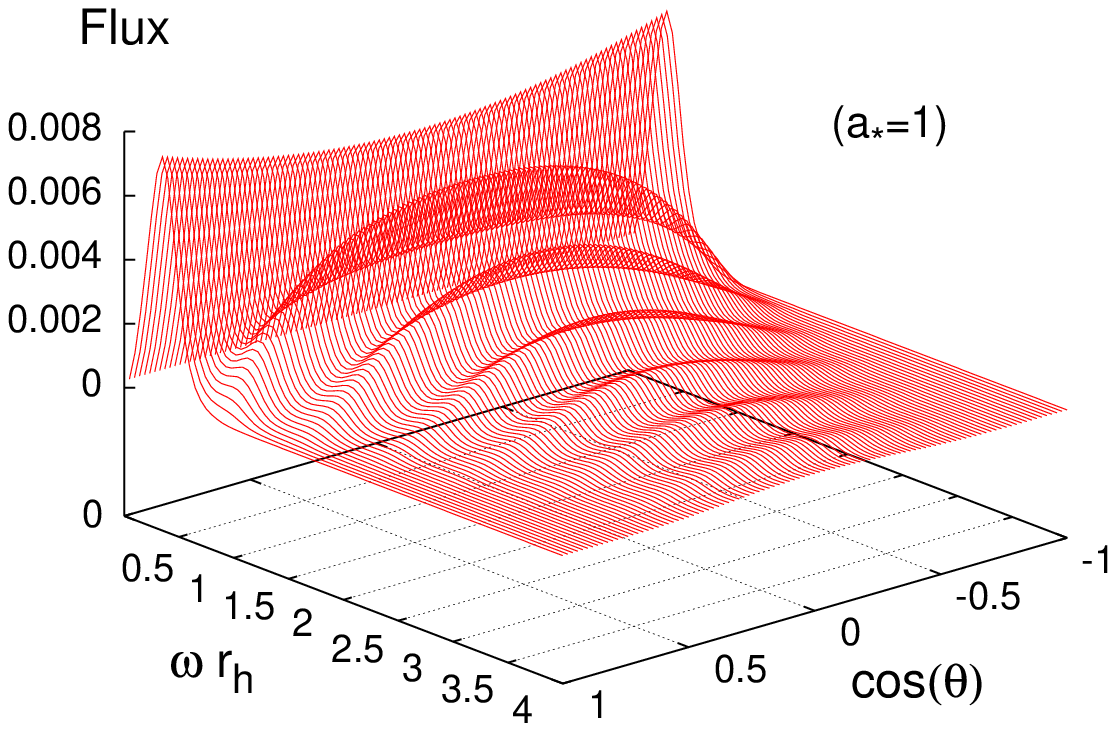, height=4.17cm}}\end{tabular}
\caption{Angular distribution of the flux spectra for emission of fermions on the brane
from a rotating black hole, for $n=1$ and $a_*=(0,0.6,1)$.}
\label{n1f-ang}
}
\FIGURE{
\begin{tabular}{c} \hspace*{-0.4cm}
\epsfig{file=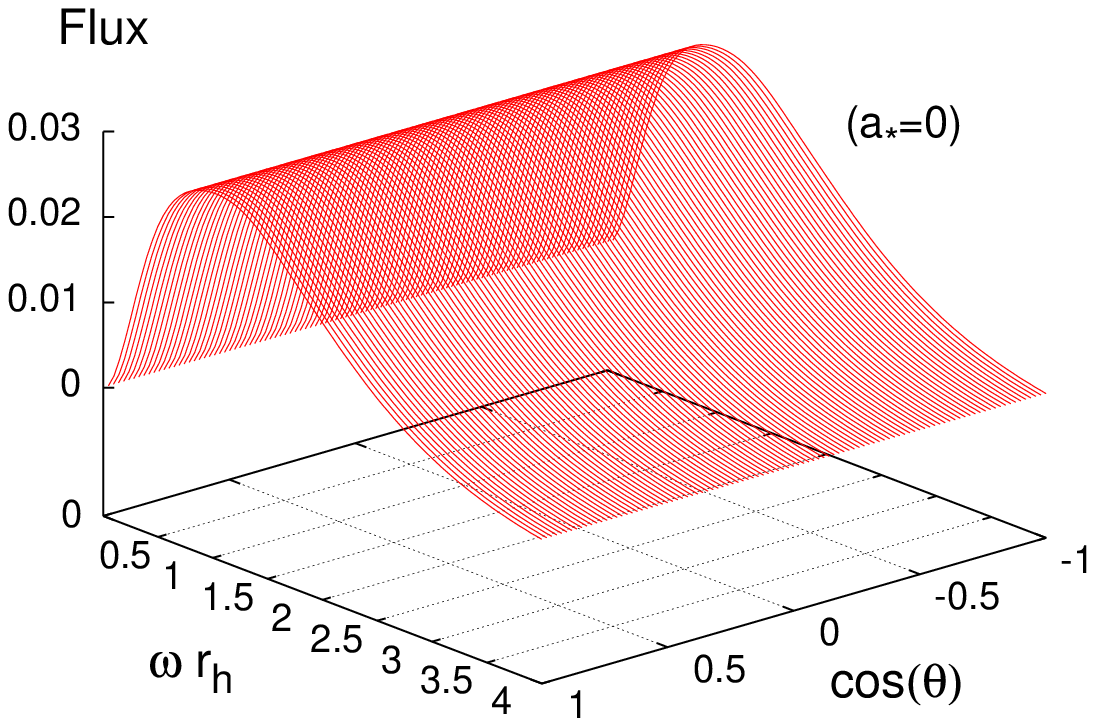, height=4.17cm}\hspace*{-1.15cm}
\epsfig{file=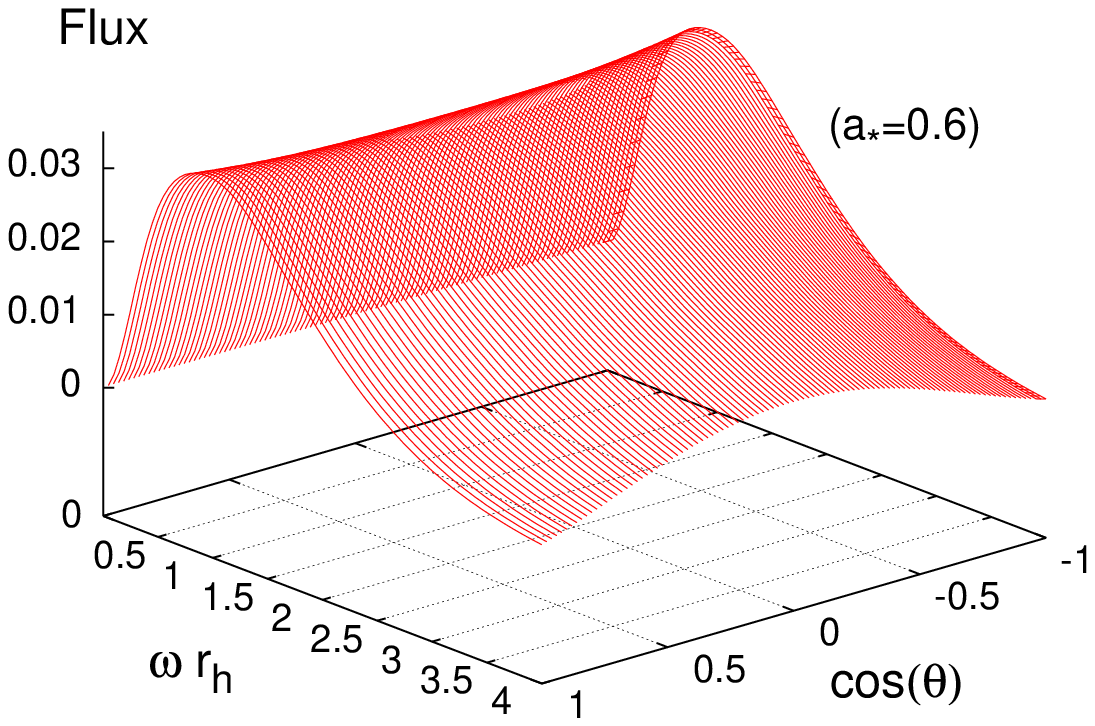, height=4.17cm}\hspace*{-1.15cm}{
\epsfig{file=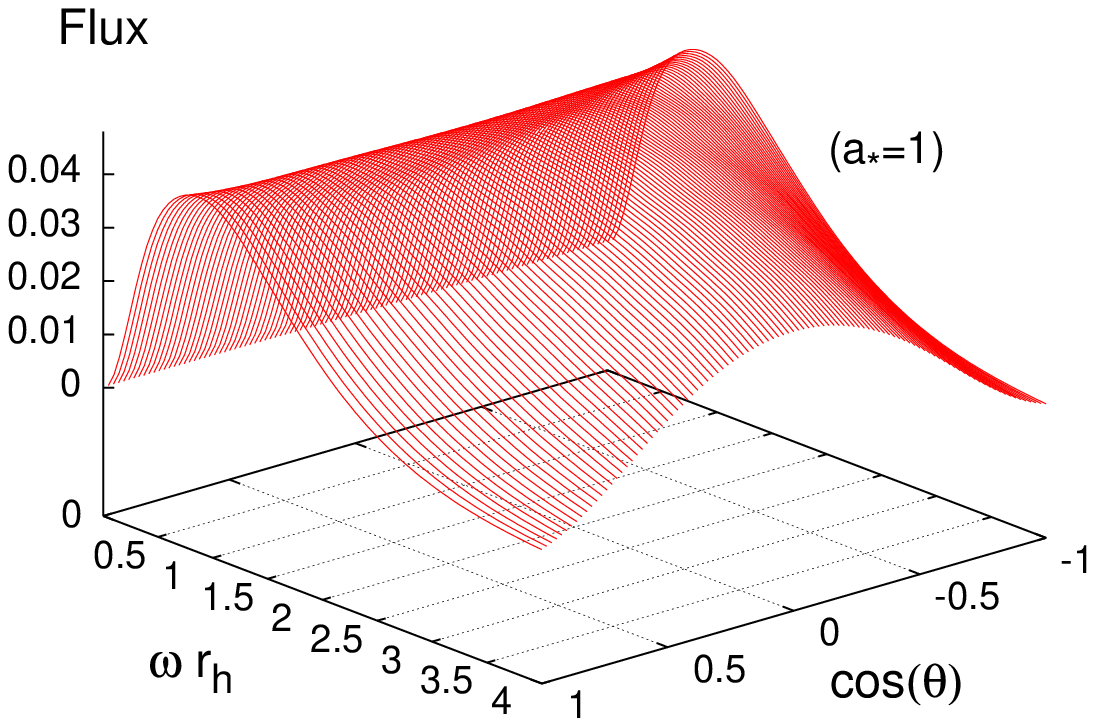, height=4.17cm}}\end{tabular}
\caption{Angular distribution of the flux spectra for
emission of fermions on the brane
from a rotating black hole, for $n=6$ and $a_*=(0,0.6,1)$.}
\label{n6f-ang}
}



\subsection{Power emission spectra}

We now turn to the power spectrum, i.e. the energy emitted by the black hole
in the form of fermions per unit time and unit frequency. Our results are
displayed in Figs. \ref{n1power} and \ref{a1power}, again for fixed
dimension and angular momentum parameter, respectively.
For fixed $n$,
as the angular momentum of the black hole increases, the energy emission curve
steadily moves towards larger values of the energy parameter due to the
increasingly larger number of high-energy particles emitted, as was observed
in the previous subsection.

As a result, significantly larger amounts of energy
are radiated away in the high-energy channels than in the low-energy ones
compared to the case of a non-rotating black hole. At the same time, each curve
becomes taller and wider, thus leading to a total energy emission rate that
is substantially enhanced. According to Fig. \ref{a1power}, a similar, but
significantly more important, enhancement is observed when the dimension
of spacetime increases while the angular momentum is kept fixed: a mere
comparison of the area outlined by the curves in the two extreme cases of
$n=1$ and $n=7$ suffices to reveal the enhancement in the emission rate,
as the number of the transverse-to-the-brane dimensions
increases\,\footnote{A comparison between our results and the ones produced in
\cite{IOP3} for the emission of fermions reveals a discrepancy of 3-5\%. This
might be due to either the failure of the analytic approximation to the angular
eigenvalue used in \cite{IOP3}, the potentially different number of partial
modes included in the calculation, or simply the different methodology used to
determine the transmission coefficient.}.

\FIGURE{
\includegraphics[height=5cm,clip]{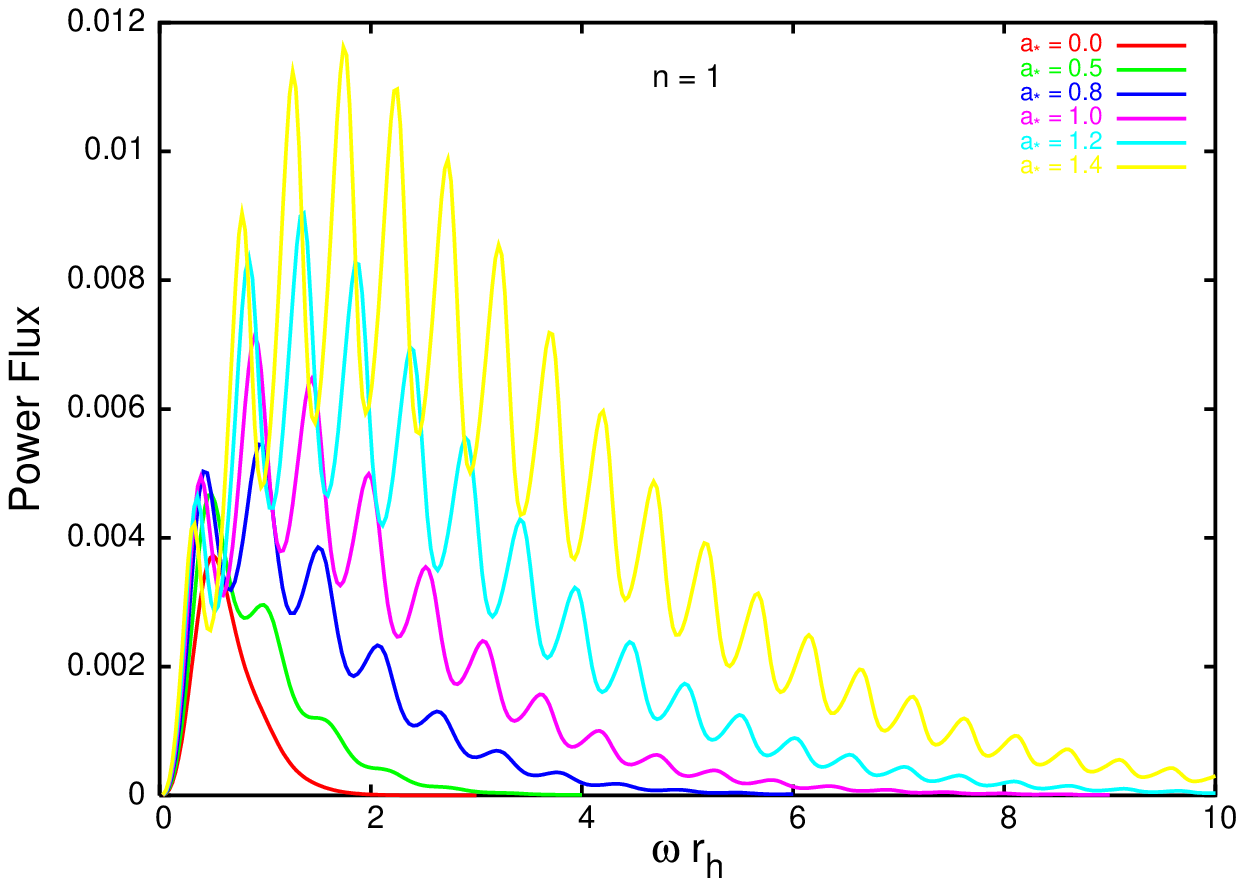}
\includegraphics[height=5cm,clip]{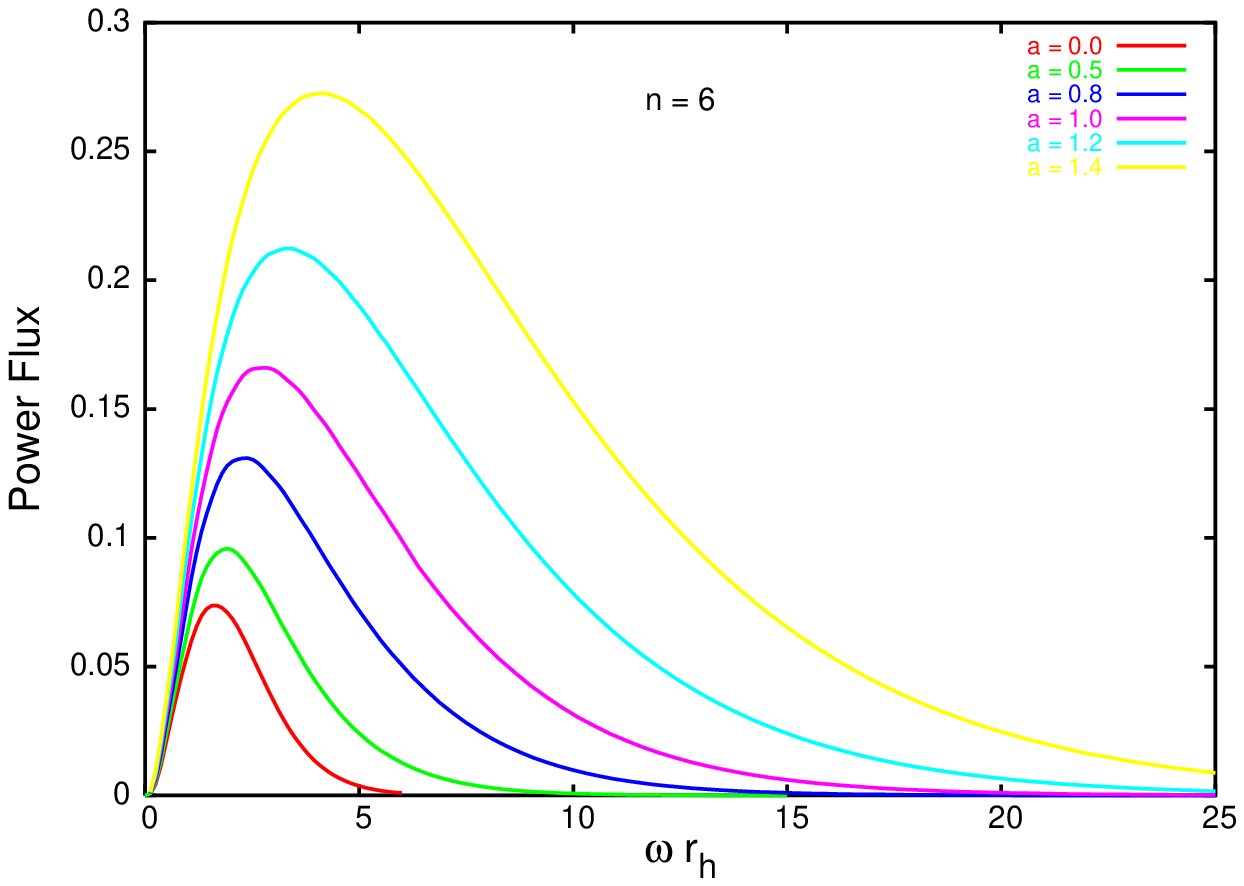}
\caption{Power emission spectra for spin-1/2 particles on the brane from a rotating black hole,
for {\bf {(a)}} $n=1$, and {\bf {(b)}} $n=6$, and various values of $a_*$.\hspace*{1.5cm}}
\label{n1power}
}
\FIGURE{
\parbox{12cm}{\begin{center}
\includegraphics[height=6cm,clip]{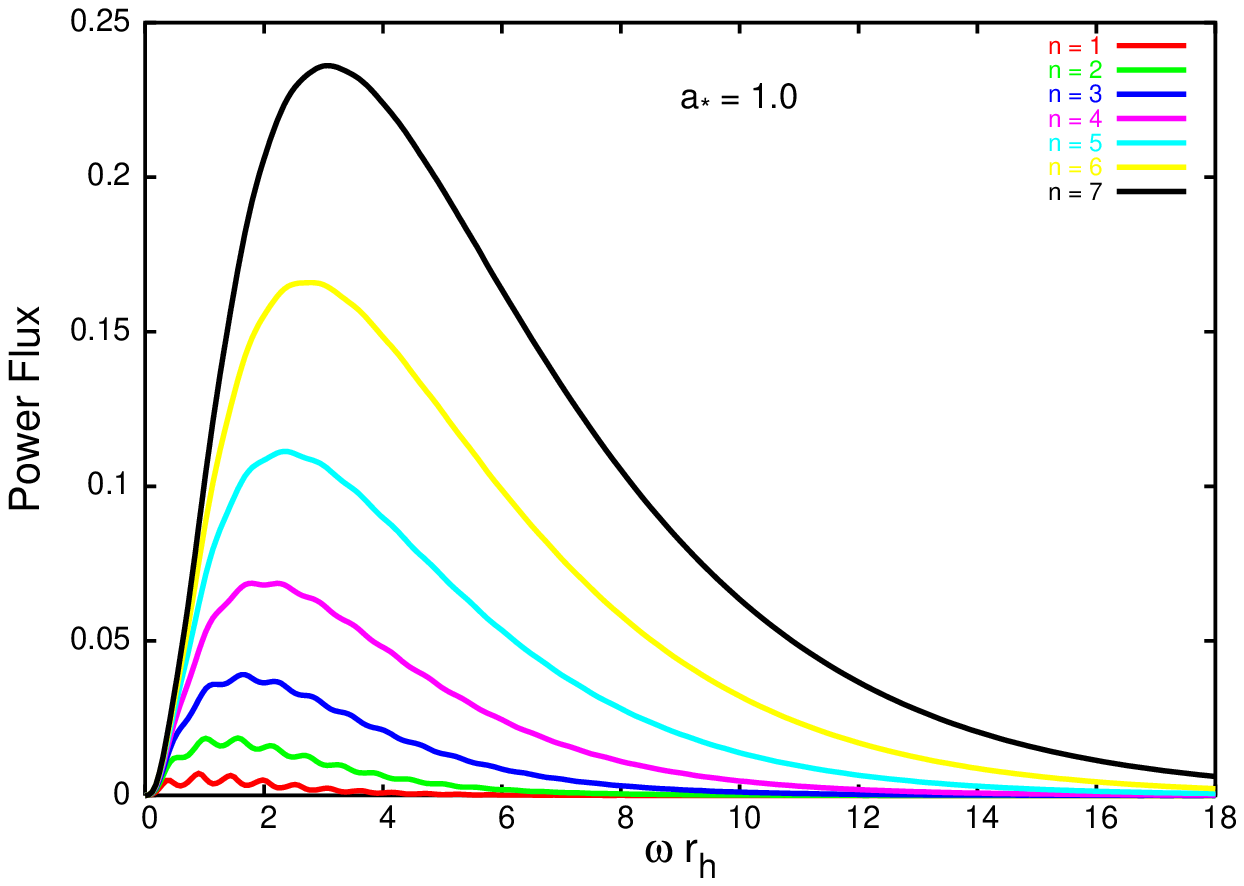}
\caption{Power emission spectra for spin-1/2 particles on the brane from a rotating black hole,
with $a_*=1$, for various values of $n$.\hspace*{1.5cm}}
\label{a1power}
\end{center}}
}

The angular distribution of the power spectra for fermion emission on the brane
was investigated next, and the results are shown in Figs. \ref{n1pow-ang} and
\ref{n6pow-ang}, for the indicative cases again of $n=1$ and $n=6$, respectively,
and for $a_*=(0,0.6,1)$.
For $a_*=0$, the spectrum has no angular dependence
as expected, however, as soon as the angular momentum of the black hole is
turned on, the effects of the centrifugal force and the spin-rotation
coupling become also apparent here.
The latter factor features in the angular
distribution pattern only in the low-energy regime and only in the case of
low dimension ($n=1$) -- as discussed earlier, as either $n$ or $a_*$
increases, the low-energy part of the spectrum becomes increasingly more
suppressed compared to the high-energy one which results also in the
disappearance of the polarization of the radiation along the rotation axis.
As in the case of the flux spectra, the centrifugal force prevails in the
high-energy emission channels, and the bulk of the energy is clearly emitted
along the equatorial plane.
\FIGURE{
\begin{tabular}{c} \hspace*{-0.4cm}
\epsfig{file=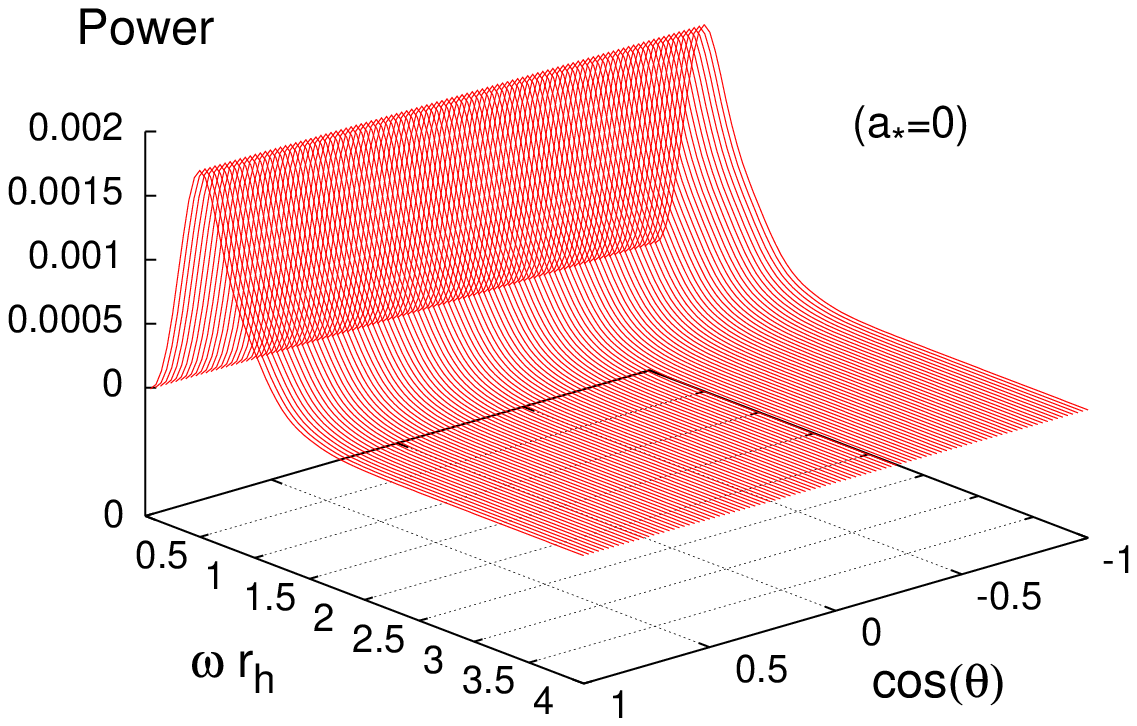, height=4.17cm}\hspace*{-1.15cm}
\epsfig{file=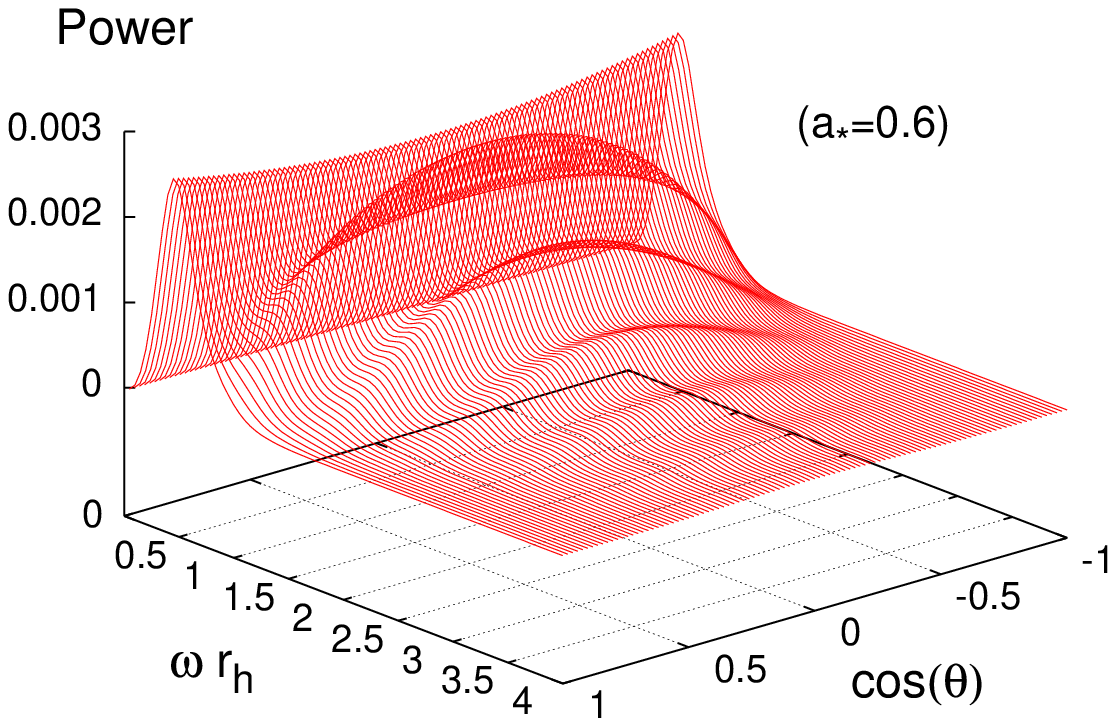, height=4.17cm}\hspace*{-1.15cm}{
\epsfig{file=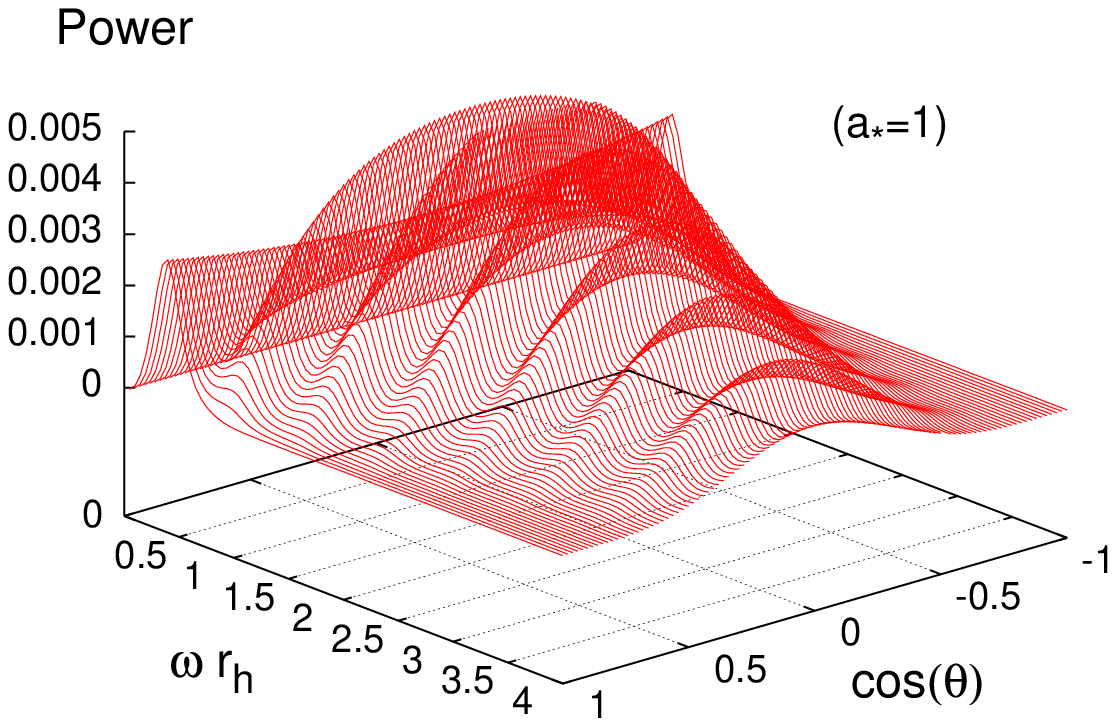, height=4.17cm}}\end{tabular}
\caption{Angular distribution of the power spectra for emission of fermions on the brane
from a rotating black hole, for $n=1$ and $a_*=(0,0.6,1)$.}
\label{n1pow-ang}
}
\FIGURE{
\begin{tabular}{c} \hspace*{-0.4cm}
\epsfig{file=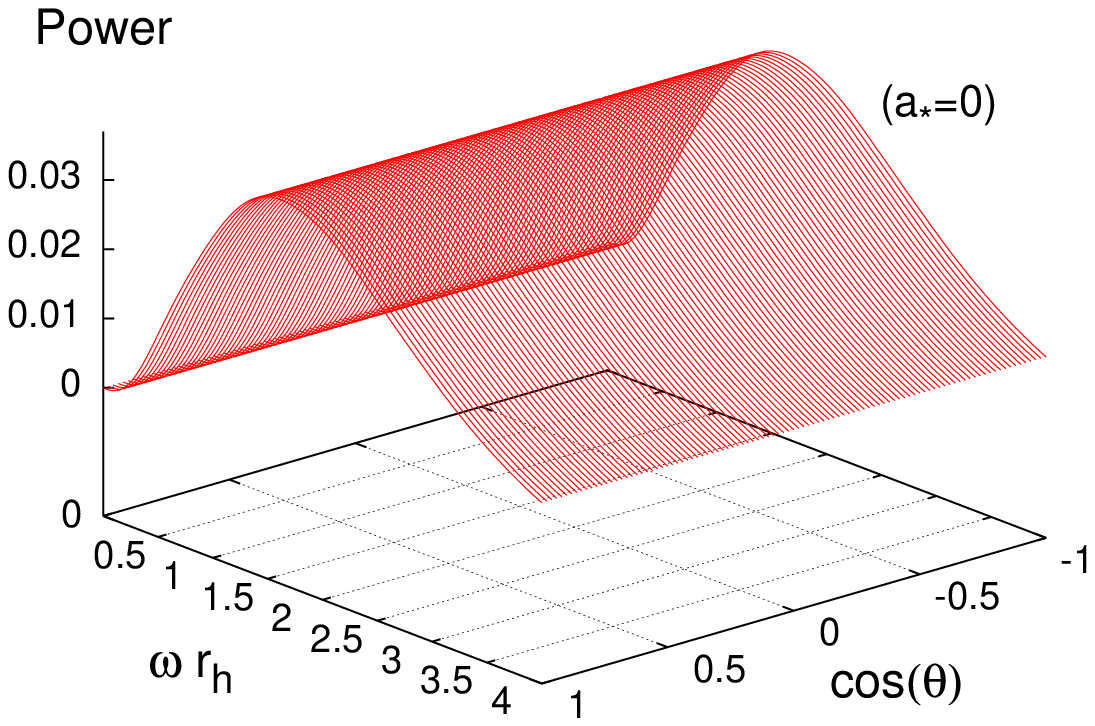, height=4.17cm}\hspace*{-1.15cm}
\epsfig{file=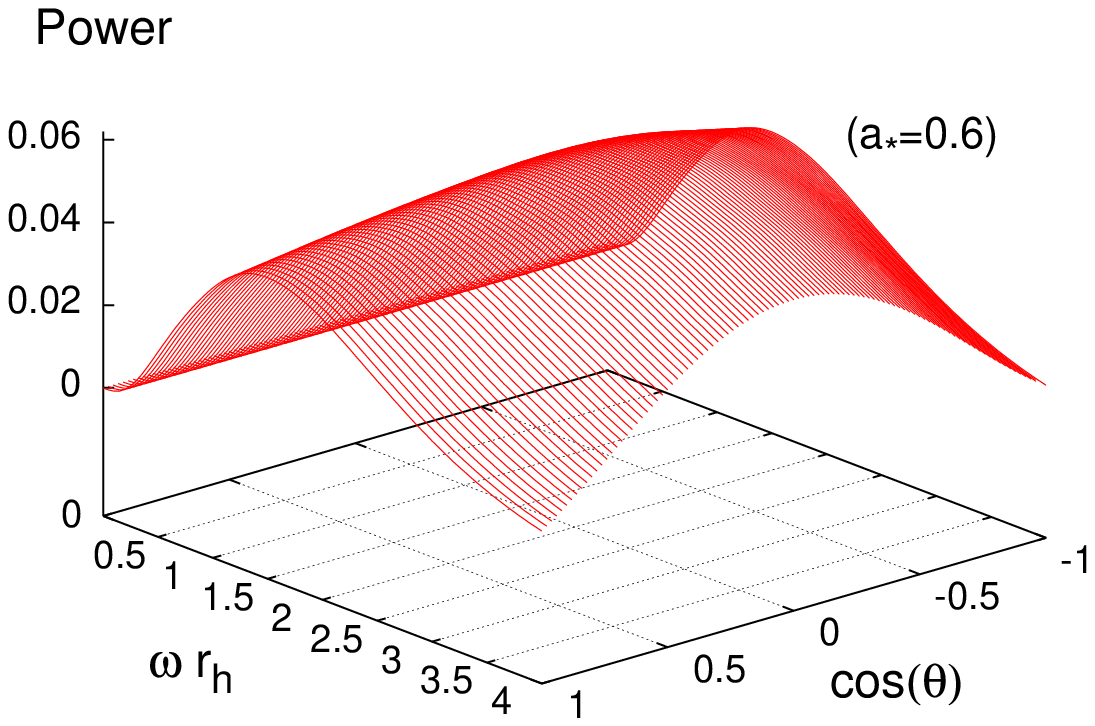, height=4.17cm}\hspace*{-1.15cm}{
\epsfig{file=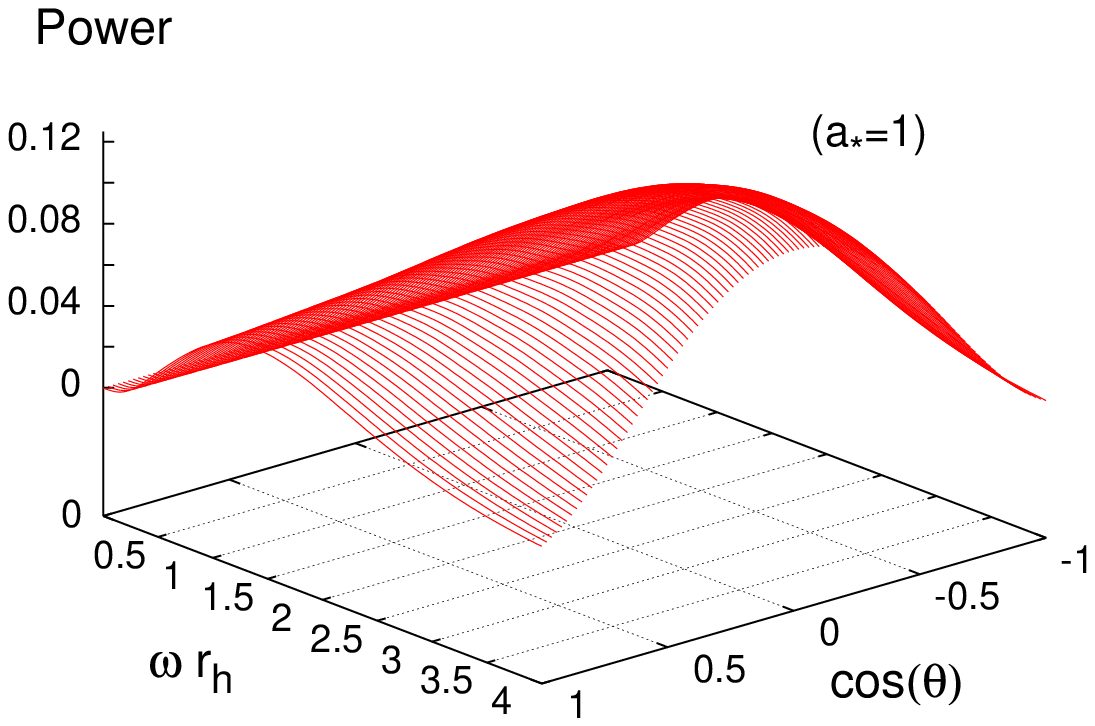, height=4.17cm}}\end{tabular}
\caption{Angular distribution of the power spectra for emission of fermions on the brane
from a rotating black hole, for $n=6$ and $a_*=(0,0.6,1)$.}
\label{n6pow-ang}
}
%

\subsection{Angular momentum spectra}

We now proceed to study the rate of emission of the angular momentum
of the black hole, and its dependence on $a_*$ and $n$. The amount of
angular momentum radiated by the black hole per unit time and frequency,
is shown in Figs. \ref{n1am}(a,b) for $n=1$ and $n=6$, respectively,
and variable $a_*$, and in Fig. \ref{a1ang} for fixed $a_*$ and
variable $n$.
\FIGURE{
\includegraphics[height=5cm,clip]{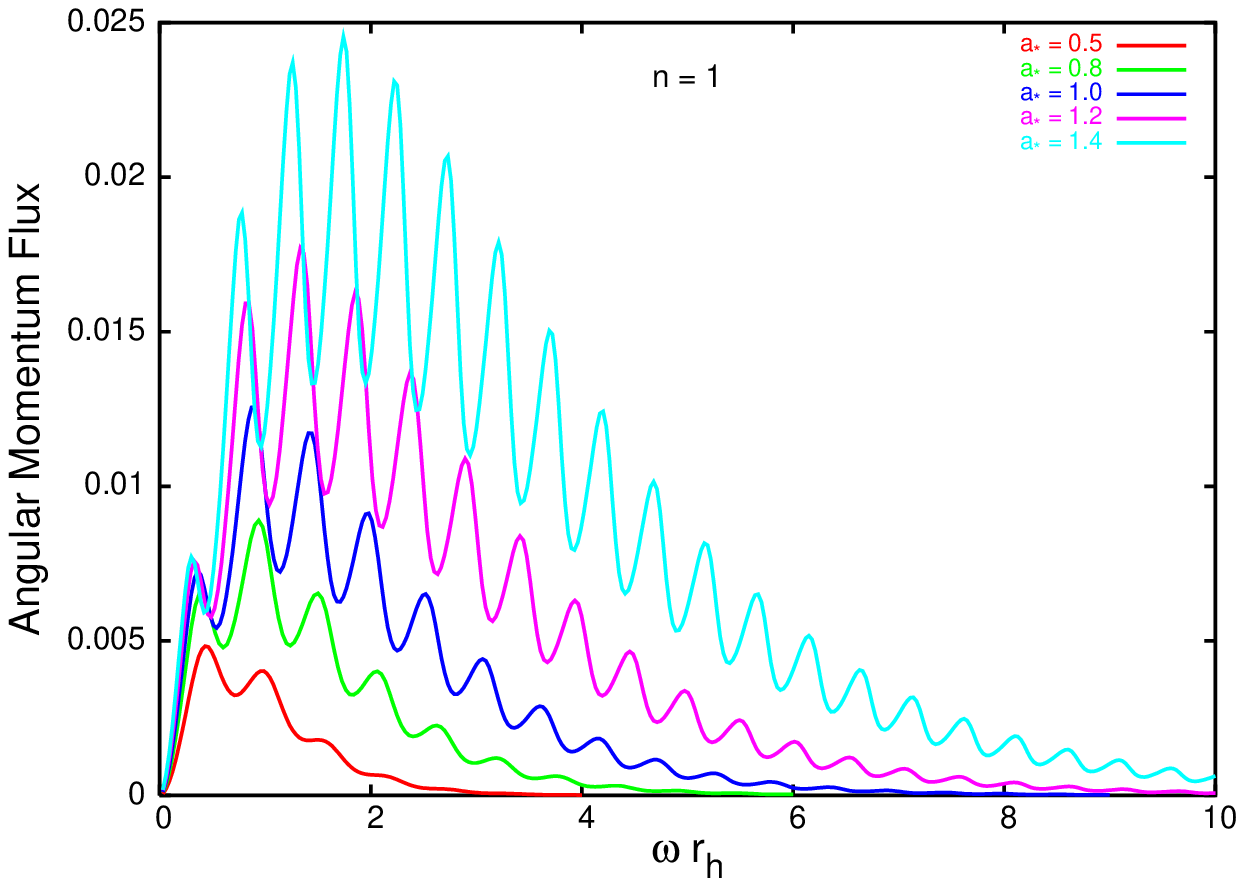}
\includegraphics[height=5cm,clip]{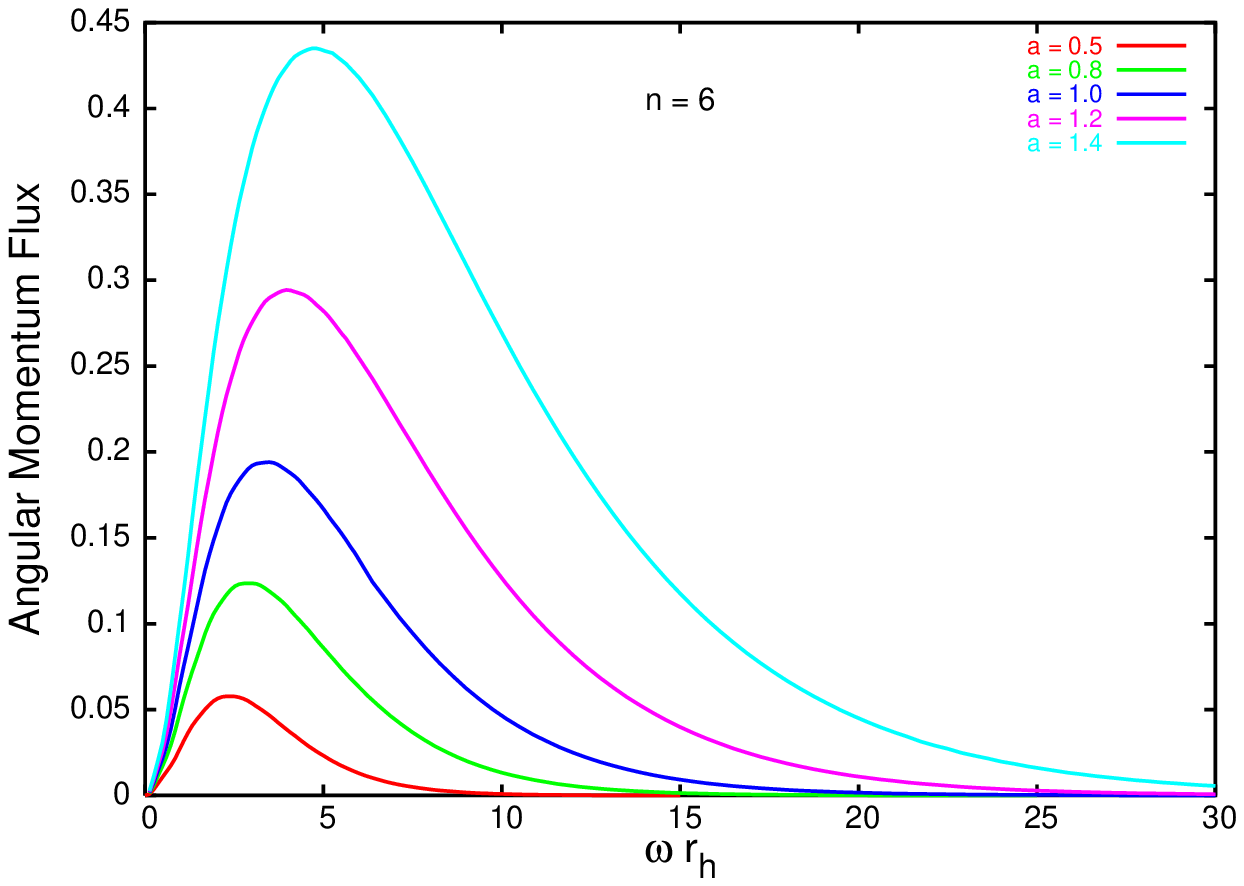}
\caption{Angular momentum spectra for spin-1/2 particles from a rotating black hole,
for {\bf {(a)}} $n=1$, and {\bf {(b)}} $n=6$, and various values of $a_*$.\hspace*{1.5cm}}
\label{n1am}
}
\FIGURE{\parbox{12cm}{\begin{center}
\includegraphics[height=6cm,clip]{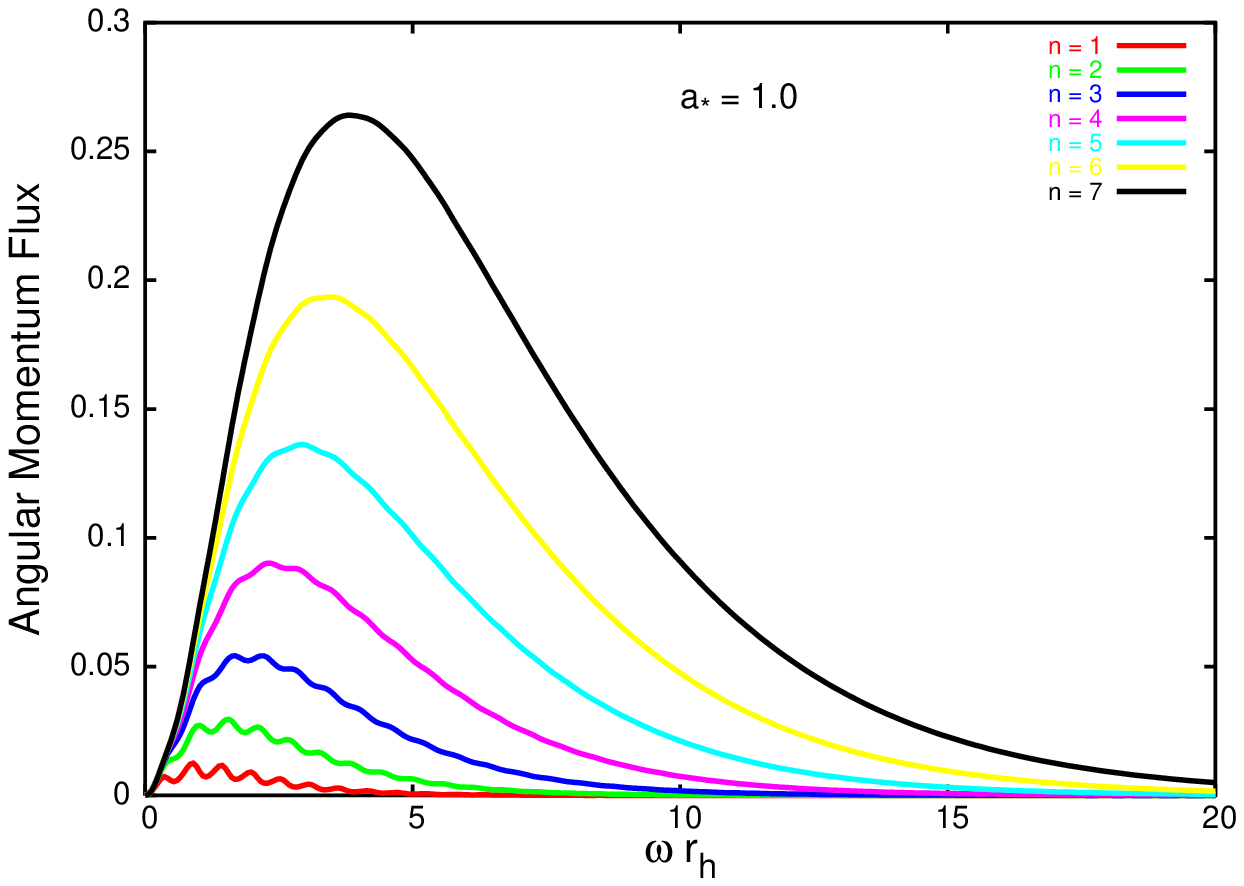}
\caption{Angular momentum spectra for spin-1/2 particles from a rotating black hole,
for $a_*=1$ and various values of $n$.\hspace*{1.5cm}}
\label{a1ang}
\end{center}}
}

As in the case of scalars and gauge bosons, the angular
momentum spectrum strongly resembles the power spectrum. As $a_*$
increases, while $n$ remains fixed, the emission curve shifts to the
right thus allowing for the loss of the angular momentum of the
black hole through the emission of higher energy particles. As either
$a_*$ or $n$ increases, the emission curve becomes again wider and taller,
thus revealing the fact that rapidly rotating black holes, or black
holes living in a spacetime with a large number of transverse dimensions
lose their angular momentum much faster.

\subsection{Total emissivities}

We finally turn to the total emissivities, that describe the emission
of either particles, energy or angular momentum per unit time, over the
whole frequency band. These total rates follow by integrating the differential
emission rates (\ref{flux})-(\ref{ang-mom}) over the energy parameter
$\omega r_h$. Here, this integration has been performed from zero to a
maximum value of the energy parameter, appropriately chosen to cover all
the emission for the cases studied. For all three types of spectra, that
maximum value was found to be $\omega r_h=30$. As a result, the computed
total emissivities are accurate, and can provide exact enhancement factors
for all three types of spectra in terms of $a_*$ and $n$.

\FIGURE{
\mbox{\includegraphics[height=5cm,clip]{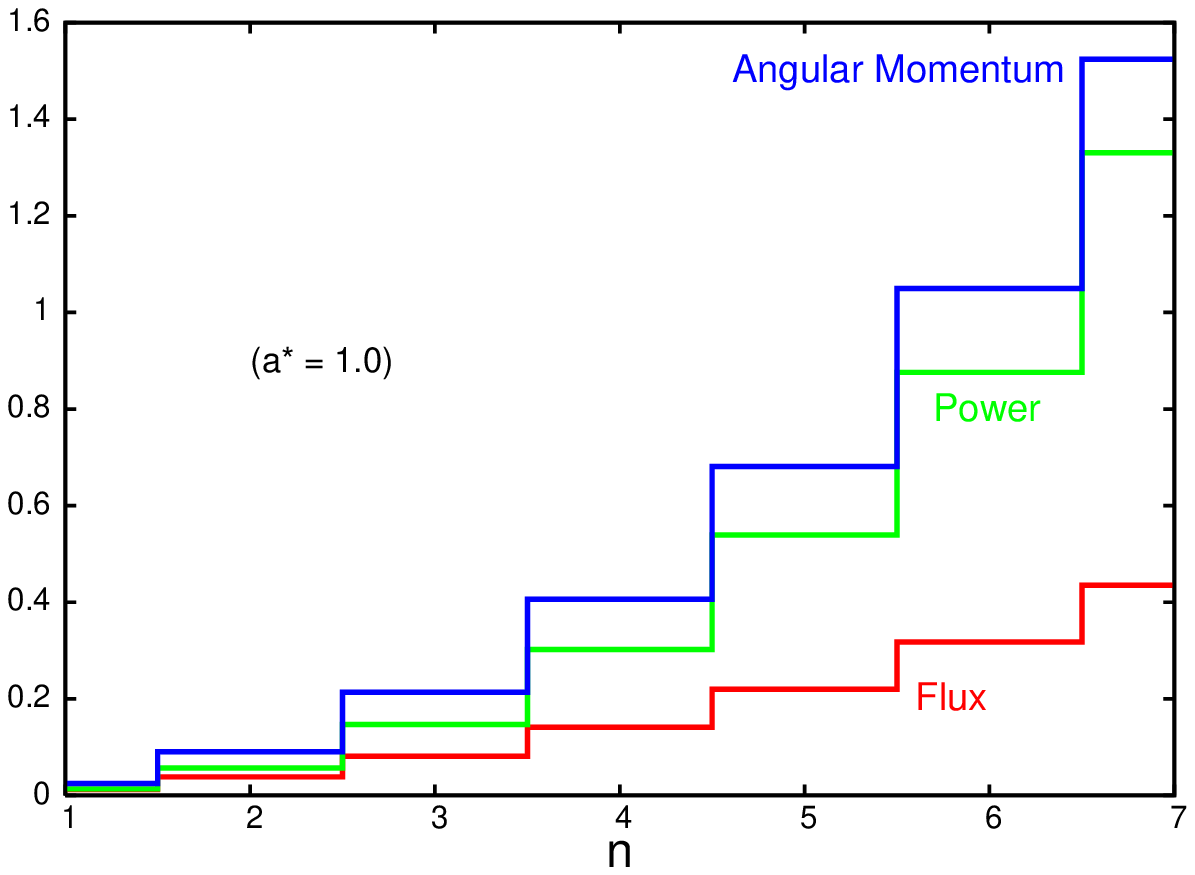}}
{\includegraphics[height=5cm,clip]{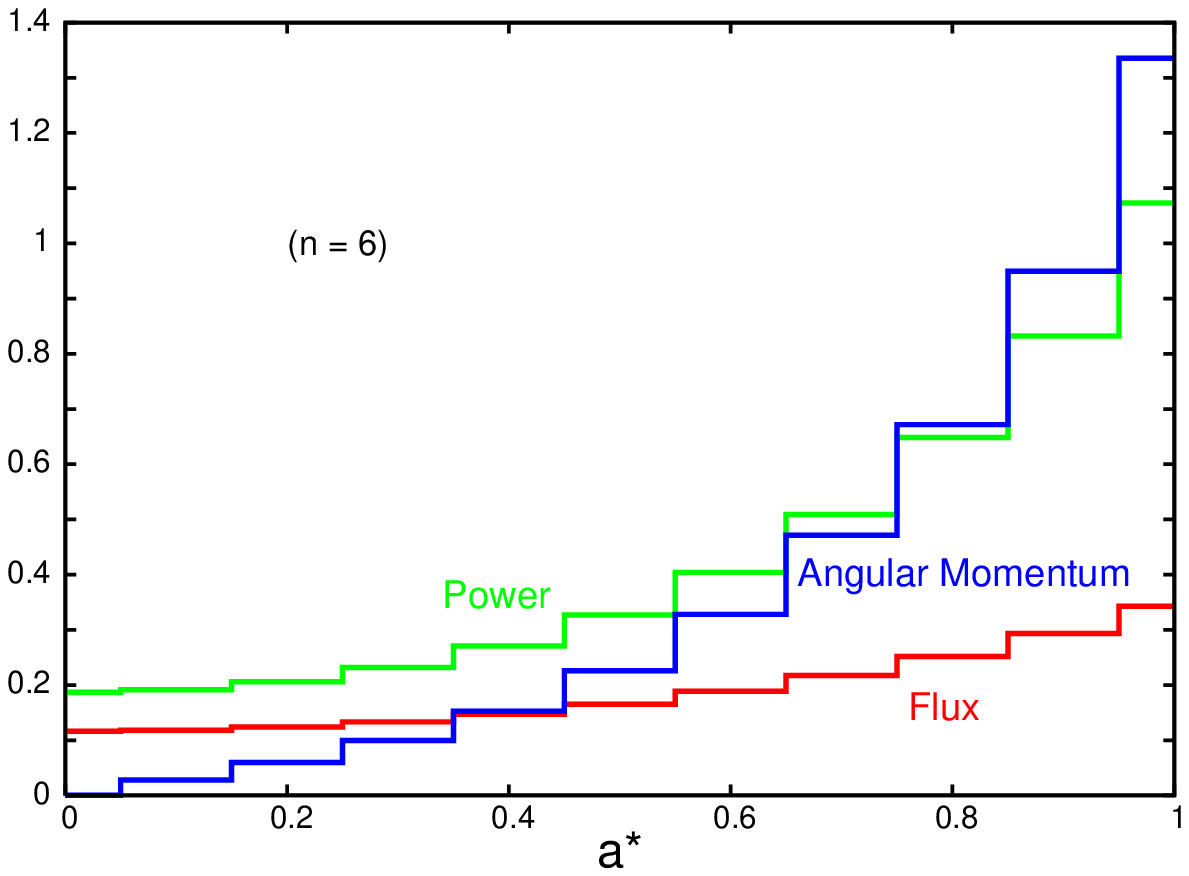}}
\caption{Total emissivities for spin-1/2 particle emission on the brane
from a rotating black hole as a function of {\bf {(a)}} $n$, for $a_*=1$, and
{\bf {(b)}} $a_*$, for $n=6$.}
\label{totalna}
}

In Figs. \ref{totalna}(a,b), we present the total emissivities of all
three types of emission rates, particle flux, energy
and angular momentum, as a function of the number of extra dimensions $n$
and angular momentum parameter $a_*$, respectively. The first histogram
reveals the enhancement of all three rates, as $n$ increases, while $a_*$
remains fixed ($a_*=1$). According to our numerical data, as $n$ goes
from 1 to 7, the total number of fermions emitted by the black hole per
unit time increases by a factor of 35.9, the total energy by a factor
of 98.7, and the total angular momentum by a factor of 61.5.
The second histogram, Fig. \ref{totalna}(b), gives the total emissivities
when the dimension of spacetime is kept fixed ($n=6$) and $a_*$
varies. As $a_*$ increases from zero to unity, a significant enhancement
takes place again, although of a smaller magnitude: the number of particles
emitted per unit time is tripled, while the energy emissivity is enhanced by
a factor of 5.7.

Let us finally discuss the angular distribution of energy emitted by the black
hole, in the form of fermions, as given by Eq. (\ref{powerang}),
but integrated again over the frequency parameter. The derived power emissivity
is then a function only of the azimuthal angle $\theta $, and for fixed $n$ and
$a_*$, is depicted in Figs. \ref{totalna-theta}(a,b), respectively.
\FIGURE{
\mbox{\includegraphics[height=5cm,clip]{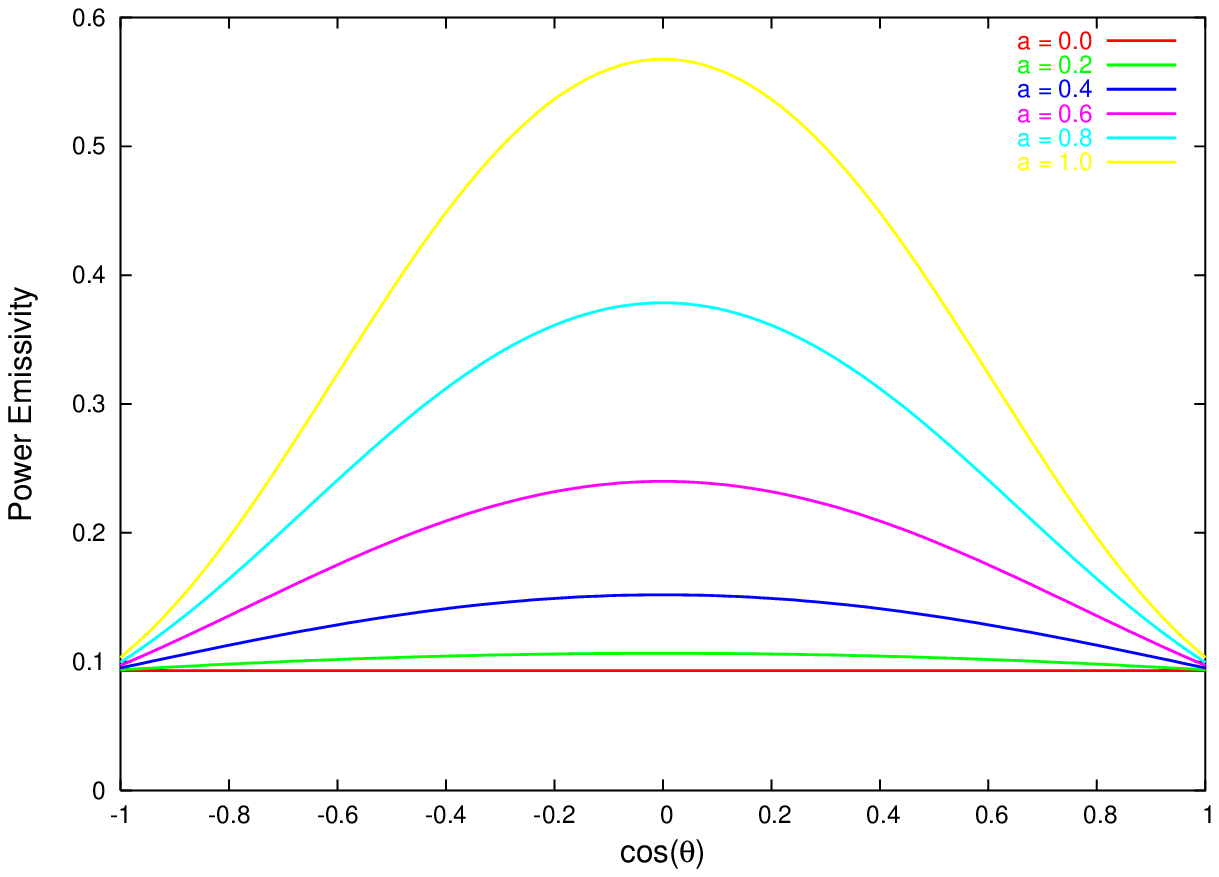}}
{\includegraphics[height=5cm,clip]{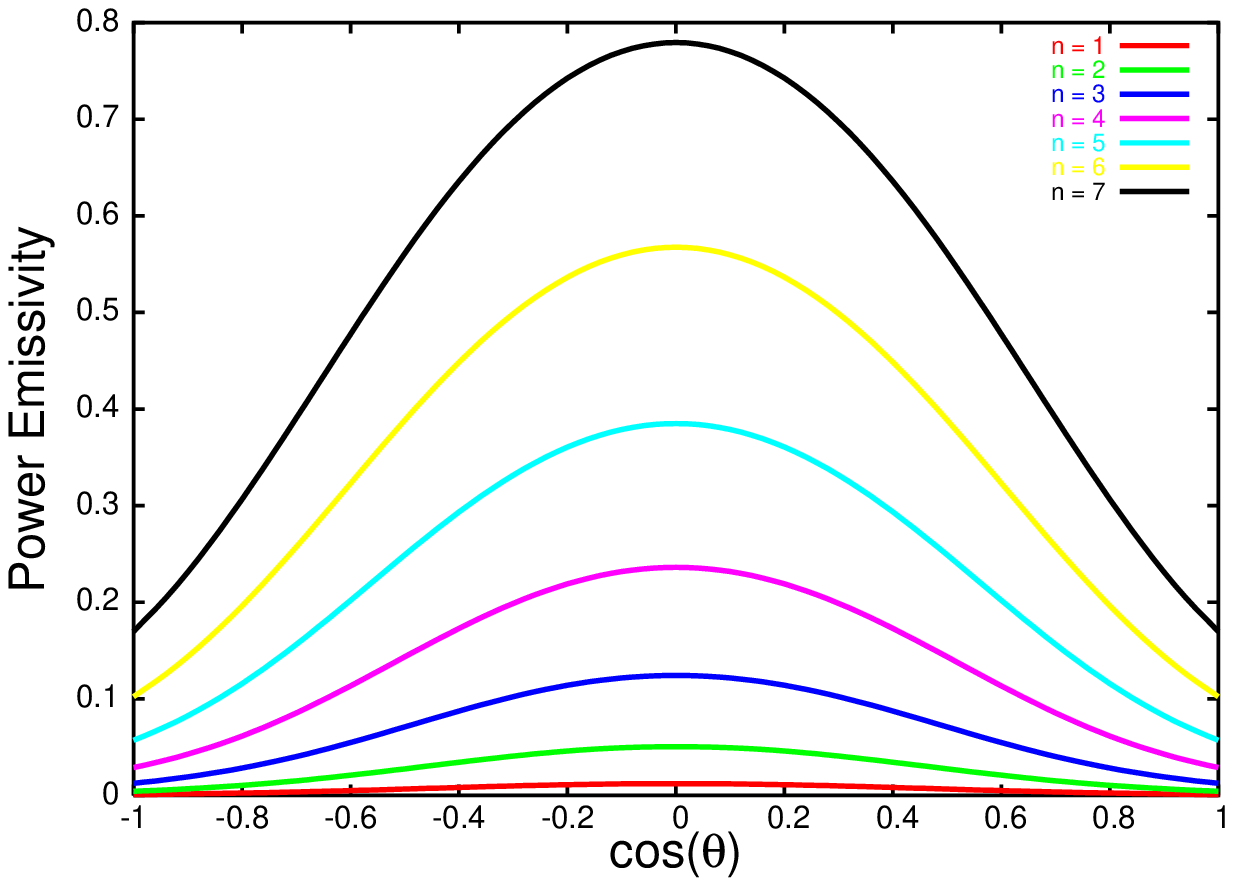}}
\caption{Power emissivity (up to $\omega r_{h}=6$) for spin-1/2 emission on the brane from a rotating
black hole as a function of $\cos \theta$, {\bf {(a)}} for $n=6$ and variable $a_*$,
and {\bf {(b)}} for $a_*=1$ and variable $n$.}
\label{totalna-theta}
}
Similarly to
the non-integrated one studied in a previous section, the power emissivity is
symmetric with respect to the equatorial plane ($\cos\theta=0$). The effect
of the centrifugal force on the angular distribution of the radiation is also
obvious here with the bulk of emitted energy being concentrated
on the equatorial plane, for all values of the angular momentum parameter and
dimension of spacetime. However, the expected polarisation along the rotation
axis caused by the spin-rotation coupling, that gave rise to a twin-peak pattern
in a similar plot for gauge bosons \cite{CKW}, is not visible here. This is due
to the fact that the low-energy regime, where the spin-rotation coupling is
effective, is significantly suppressed when the total emissivities are computed;
in addition, the magnitude of this effect for fermions was much smaller in the
first place, compared to the ones for gauge bosons, due to the smaller value of
their spin. The shape of the fermion power emissivity curves appearing in Figs.
\ref{totalna-theta} are found to agree with the ones derived in the case of a
4-dimensional, rotating black hole \cite{Leahy}.


\section{Conclusions}

In this work, we have completed the study of the Hawking radiation
on the brane from a higher-dimensional rotating black hole by
investigating the emission of fermionic modes. A comprehensive
analysis was performed that led to the particle, power and angular
momentum emission rates, and shed light on their dependence on
fundamental parameters of the theory such as the spacetime
dimension and angular momentum of the black hole. In addition, the
angular distribution of the emitted modes, in terms of number of
particles and energy, was thoroughly studied. Our results are
valid for arbitrary values of the energy
 of the emitted particles\footnote{Using the results in \cite{NS}, we have investigated at
what value of $\omega r_h$ is the contribution to the spectrum
from ultra-short distances (where a full theory of quantum gravity
would be required) of the same order as the total spectrum itself.
For the main cases plotted in this paper, such value of $\omega
r_h$ is at the tail-end of the curve for each dominant mode, so
that the main contribution to the spectrum is indeed from the
semiclassical regime.}
  and angular momentum of the black hole, and complement previous results
on the emission of brane-localised scalars \cite{HK2,DHKW} and gauge
bosons \cite{CKW}.

As in the case of our previous analyses for spin-0 and spin-1 particles,
the equation of motion for fermions propagating in the projected-on-the-brane
gravitational background needed to be determined first, a task that was
performed by employing the Newman-Penrose formalism. The derivation of
exact results for the various emission rates demanded the numerical
integration of both the angular and radial part of the equation of
motion in exactly that order. The former calculation offered exact results
for both the angular eigenfunctions, the spin-weighted spheroidal harmonics,
and the angular eigenvalues, that also appear in the radial equation.
The latter calculation led to exact results for the radial part of the
wavefunction of the fermionic field, and eventually for the transmission
coefficient.

Our numerical results for the transmission probability as well as for the
particle, energy and angular momentum fluxes were presented in Section
\ref{num-results}. The differential emission rates per unit time and frequency,
integrated over the azimuthal angle $\theta$, were addressed first, and
spectra for different values of the dimension of spacetime and the
angular momentum of the black hole were constructed as a function of the
energy of the emitted particle. As in the case of scalars and gauge bosons,
it was revealed that any increase in either $n$ or $a_*$ significantly
enhances all emission rates of the black hole. The enhancement is
present in all frequency regimes, and significantly boosts the emission of
higher energy particles compared to the case of a non-rotating black hole.
Total emissivities, i.e. emission rates integrated over all frequency regimes,
were also presented at the end of Section \ref{num-results}; these total
emission rates confirmed the enhancement of all spectra, in terms of $n$
and $a_*$, and led to the derivation of the corresponding enhancement factors:
as $n$ increases, the various emissivities are enhanced by a factor that varies
from 35 to almost 100, while, as $a_*$ grows, this factor is of the order
of 3 or 6.

Throughout the study of the emission of various spin-$s$ particles on
the brane by a higher-dimensional rotating black hole, a significant amount
of energy was invested in the calculation of the exact spin-weighted spheroidal
harmonics, that would allow us to determine the exact angular distribution of the
emitted particles and energy. Contrary to the case of a spherically-symmetric
black hole, the various emission rates for a rotating black hole bear a
signature of a characteristic direction in space which is the axis of rotation.
It was found that the majority of particles emitted by the black hole are
restricted to do so along the equatorial plane due to the strong centrifugal
force - this conclusion was reinforced after the total energy emissivities
as a function of the azimuthal angle were also computed. Nevertheless, the
spin of the particle can significantly affect the angular distribution pattern
at the low-energy regime due to the spin-rotation coupling with the angular
momentum of the black hole.
For comparison, in Figs. \ref{comparison1}(a,b,c) we present the
angular distribution of the energy emission spectrum for a 6-dimensional
($n=2$) rotating black hole emitting radiation on the brane, in the
form of scalars \cite{DHKW}, fermions and gauge bosons \cite{CKW},
respectively. From these, it becomes obvious that the angular variation
in the pattern of the emitted radiation can be a distinctive signature
of emission from a rotating black hole, but also of the spin and energy
of the particles emitted.

\FIGURE{
\begin{tabular}{c} \hspace*{-0.4cm}
\epsfig{file=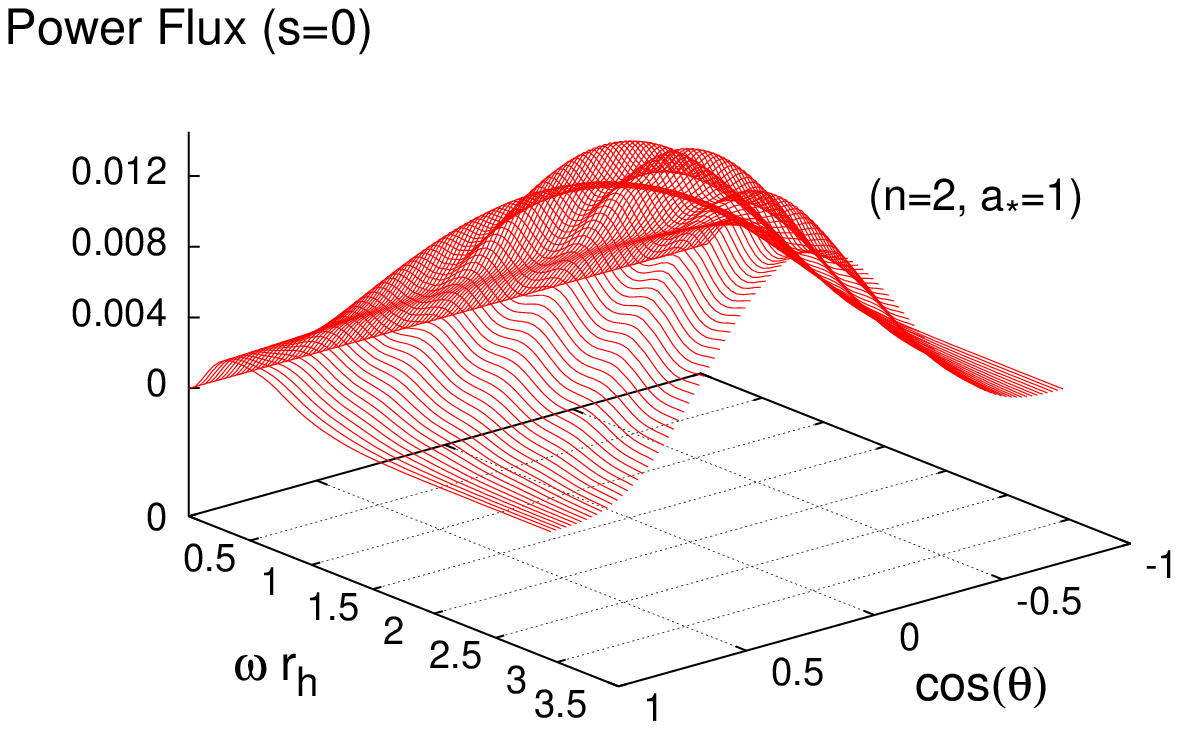, height=4cm}\hspace*{-1.25cm}
\epsfig{file=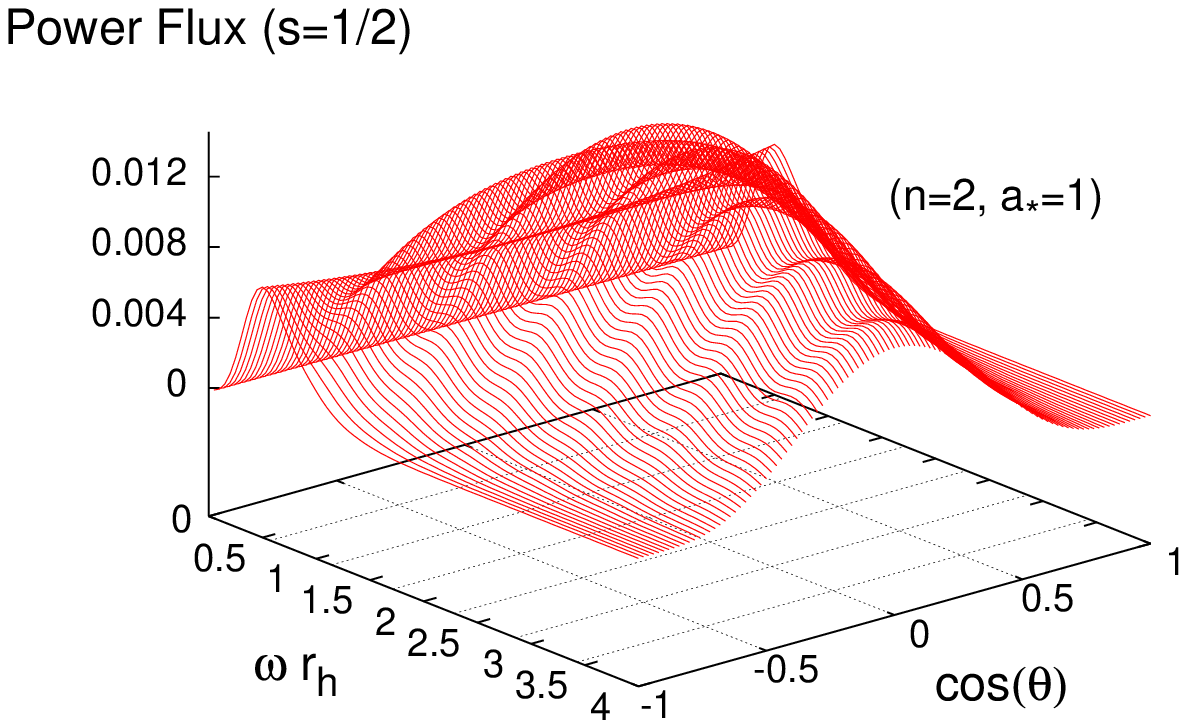, height=4cm}\hspace*{-1cm}{
\epsfig{file=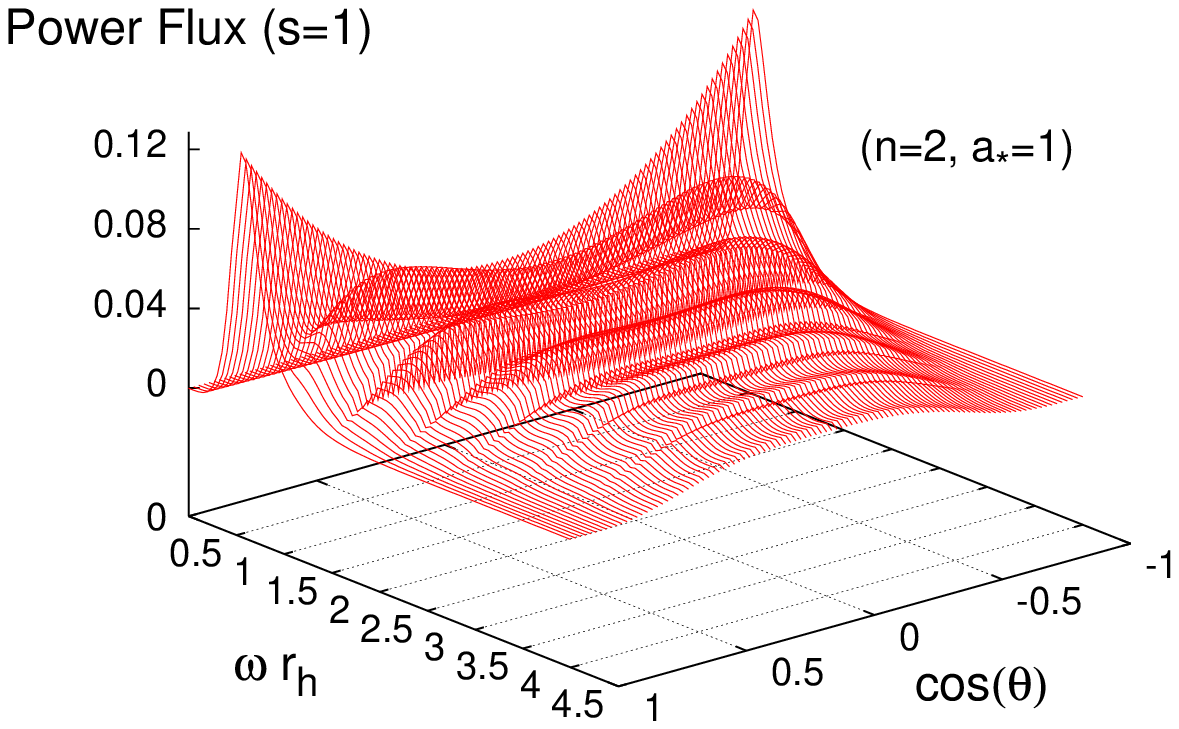, height=4cm}}\end{tabular}
\caption{Angular distribution of the power spectra for {\bf (a)} scalars,
{\bf (b)} fermions, and {\bf (c)} gauge bosons on the brane
from a 6-dimensional black hole with $a_*=1$.\label{comparison1}}
}

The array of graphs displayed in Fig. \ref{comparison1} can shed light also
on the question of the relative emissivities of different types of particles
on the brane from a rotating black hole. In the absence of angular momentum,
it was shown \cite{HK1} that the preference of the black hole for the
emission of scalar fields in four dimensions is replaced by the one for
gauge bosons when a number of additional spacelike dimensions is introduced.
In \cite{CKW}, we were able to show that a higher-dimensional rotating black
hole also prefers to emit gauge bosons over scalars on the brane. With
the results for fermions being now available, the complete picture has
emerged. By merely comparing the vertical axes of the graphs in Fig.
\ref{comparison1}, one may see that a higher-dimensional rotating black hole
emits approximately the same amount of energy on the brane in the form of
scalars and fermions -- a more careful comparison of the emission rates
integrated over the azimuthal angle, presented in Fig. \ref{n1power}, with
the corresponding ones in \cite{DHKW} reveals that the fermions slightly dominate
over the scalars. Nevertheless, as Fig. \ref{comparison1}(c) reveals, a
significantly larger amount of energy (by, at least, an order of magnitude)
is spent by the black hole in the emission of gauge bosons -- the above
result holds for all values of $n$.

The results presented in this work, addressing modifications of the properties
of black holes submerged into a higher-dimensional brane-world spacetime,
have an obvious theoretical interest. The potential detection of the
Hawking radiation from a higher-dimensional black hole formed during a
high-energy particle collision in a ground-based accelerator, in the
near or far future, increases their significance as the corresponding
spectra offer vital information about fundamental parameters of the
system under study, such as the dimension of spacetime and angular
velocity of the black hole, as well as for the creation of the black
hole itself. This information is obtained by looking only
at the emission channel along our brane where ordinary Standard Model
particles, as well as the potential observers, are restricted to live.
With the investigation of the spin-down and Schwarzschild phases of the life
of the black hole now completed, what remains to be done is a realistic
study of the dynamical evolution of the produced black hole where more
advanced aspects, such as the variation of the temperature of the black
hole as the decay progresses or the properties and interactions of the
Standard Model particles, are properly taken into account. This goal will
be achieved by entering our results for the emission of individual scalar,
fermionic and gauge bosonic degrees of freedom from a constant-mass black
hole into the CHARYBDIS black hole event generator \cite{HRW}. We hope
to report results from this analysis soon.


\acknowledgments
The work of P.K. is funded by the UK PPARC Research Grant
PPA/A/S/2002/00350. The work of E.W. is supported by UK PPARC,
grant reference number PPA/G/S/2003/00082. M.C. wishes to thank
Paul Watts for helpful discussions on the spinor formalism and
Jose Navarro-Salas and Ivan Agull\'{o} for their insight into
transplanckian effects on Hawking radiation spectra.



\end{document}